\documentclass[twocolumn,preprintnumbers,amsmath,amssymb,pra]{revtex4}

\usepackage{graphicx,amsmath}
\usepackage{dcolumn}
\usepackage{bm}
\usepackage{amssymb}
\usepackage{graphicx}
\usepackage{epstopdf}
\usepackage{color}
\graphicspath{{Figures/}}

\begin{document}

\title{Single-photon scattering on a two-qubit system. Spatio-temporal structure of the scattered field}

\begin{abstract}
In this paper, we study the spatiotemporal distribution of the
photon electric field produced by the scattering of a single
photon narrow pulse from a system of two identical qubits coupled
to continuum modes in a one-dimensional (1D) open waveguide. We
derive the time-dependent dynamical equations for qubits' and
photon amplitudes which allow the calculation of the photon
backward and forward scattering fields in the whole space: before
qubits, between qubits, and behind the qubits. The scattered field
consists of several contributions which describe a free field of
incoming photon, a spontaneous exponential decay of excited
qubits, a slowly decaying part that dies out as the inverse powers
of $t$, and a lossless part that represents a steady state
solution as $t\rightarrow\infty$. For our system, we find the
transmittance and reflectance fields as both time and distance
from the qubits tend to infinity. We show that as the time after
the event of scattering tends to infinity, the steady state photon
the field is being formed in the whole one-dimensional space.  If
the distance $d$ between qubits is equal to the integer of the
wavelength $\lambda$, the field energy exhibits temporal beatings
between the qubit frequency $\Omega$ and the photon frequency
$\omega_S$ with the period $T=2\pi/(\omega_S-\Omega)$.

\end{abstract}


\date{\today}

\author{Ya. S. Greenberg}\email{yakovgreenberg@yahoo.com}
\affiliation{Novosibirsk State Technical University, Novosibirsk,
Russia}

\author{A. A. Shtygashev} \affiliation{Novosibirsk State
Technical University, Novosibirsk, Russia}

\author{A. G. Moiseev} \affiliation{Novosibirsk State
Technical University, Novosibirsk, Russia}

\maketitle

\section{Introduction}

Manipulating the propagation of photons in a one-dimensional
waveguide coupled to an array of two-level atoms (qubits) may have
important applications in quantum devices and quantum information
technologies \cite{Rai2001, Roy2017, Sher2023}.

Among a variety of quantum systems which have been proposed for
the implementation of quantum processor architecture,
superconducting qubits have emerged as one of the leading
candidates in the race to build a scalable quantum computer
capable of realizing computations beyond the reach of modern
supercomputers \cite{Gu2017, Krantz2019,Kjaer2019}.

The use of a single or few photons as a probe of an array of
two-level real or artificial atoms (qubits) embedded in a 1D open
waveguide has been extensively studied both theoretically
\cite{Ruos2017,Lal2013,Chang2012} and experimentally
\cite{Mirho2019,Brehm2021,Loo2014}.

Most of theoretical calculations of the transmitted and reflected
photon amplitudes  in a 1D open waveguide with the atoms placed
inside have been performed within a framework of the stationary
theory in a configuration space
\cite{Shen2009,Cheng2017,Fang2014,Zheng2013} or by alternative
methods such as those based on Lippmann-Schwinger scattering
theory \cite{Roy2011, Huang2013, Diaz2015}, the input-output
formalism \cite{Fan2010, Lal2013, Kii2019}, the non-Hermitian
Hamiltonian \cite{Green2015}, and the matrix methods
\cite{Green2021,Tsoi2008}.

Even though the stationary theory of the photon transport provides
a useful guide to what one would expect in real experiment, it
does not allow for a description of the dynamics of a qubit
excitation and the evolution of the scattered photon amplitudes.

In practice, the qubits are excited by the photon pulses with
finite duration and finite bandwidth. Therefore, to study the real
time evolution of the photon transport and atomic excitation the
time-dependent dynamical theories have been developed
\cite{Chen2011, Liao2015, Liao2016a, Liao2016b, Zhou2022,
Green2018}. In most of these theories the evolution of qubits'
amplitudes and the transmission and reflection coefficients have
been investigated. The evolution of the photon pulse scattered
from a single qubit was studied in \cite{Dom2002, Green2023}.

It is worth noting that the conditions for the detection of the
radiation from superconducting qubits significantly differ from
those for real atoms. For example, when detecting the resonance
fluorescence the distance between the atoms is usually small
compared to their distance from the detector. However, in the
current realization of superconducting qubits with associated
circuitry the pulse generator and readout amplifier (detector)
should be placed as close as possible to the qubit system
\cite{Lec2021} where near-field effects may have a substantial
impact on the output signal. From this point, it is important to
study the spatio-temporal distribution of the electric field of a
scattered photon from a qubit system.

\begin{figure}
  \includegraphics[width=8 cm]{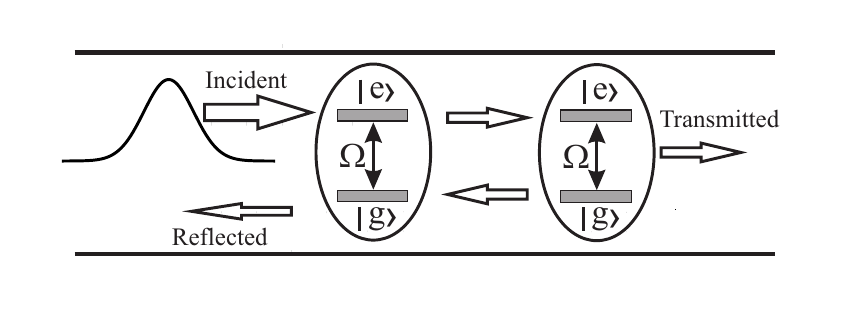}\\
  \caption{Schematic representation of a single-photon Gaussian pulse
  interacting with a two-qubit system with energy levels $|g\rangle$ and $|e\rangle$,
respectively. $\Omega$ is the separation between the energy
levels. Long horizontal lines denote the waveguide
geometry.}\label{Fig1}
\end{figure}

In this paper we study in detail the spatio-temporal distribution
of the photon electric field produced by the scattering of a
single photon narrow pulse from a system of two identical qubits
embedded in a one-dimensional open waveguide (see Fig.\ref{Fig1}).
The method we apply here is the extension of our time-dependent
theory that has been developed earlier for the case of a single
qubit \cite{Green2023}. Our method consists of two steps. First,
we rely on  the Wigner-Weisskopf approximation in which the rate
of spontaneous emission to the guided mode is much less than the
qubit frequency, $\Gamma(\omega)\ll\Omega$. In this case, a
reasonable assumption is to consider the decay rate $\Gamma$ to be
frequency independent taking it at the qubit frequency $\Omega$,
$\Gamma(\omega)=\Gamma(\Omega)\equiv\Gamma$. For a system which
consists of more than a single qubit the Wigner-Weisskopf
approximation is equivalent to Markov approximation which neglects
the retardation effects: all qubits feel the scattered field
simultaneously. This approximation allows us to obtain the
explicit expressions for the forward and backward photon
scattering amplitudes. Second, we calculate the associated
electric fields, $u(x,t)$ for forward and $v(x,t)$ for backward
travelling waves in the whole 1D space: before first qubit,
between qubits, and behind the second qubit. The fields consist of
several contributions which describe a free field of incoming
photon, a spontaneous exponential decay of excited qubits, a
slowly decaying part which dies out as the inverse powers of $t$,
and a lossless part which represents a steady state solution as
$t\rightarrow\infty$.  We show that if the distance $d$ between
qubits is equal to integer of the wavelength $\lambda$, the field
energies $|u(x,t)|^2$ and $|v(x,t)|^2$ exhibit a temporal beatings
between the qubit frequency $\Omega$ and the photon frequency
$\omega_S$. The transmittance and reflectance are also obtained
from the steady state solution for $x\rightarrow\infty$ and
$x\rightarrow-\infty$, respectively.

Even though our treatment can be applied to real two-level atoms,
we consider in our paper an artificial two-level atoms,
superconducting qubits operating at microwave frequencies at GHz
range. For our calculations we take qubits' frequency
$\Omega/2\pi=5$ GHz which corresponds to wavelength $\lambda=6$
cm.

The paper is organized as follows. In Section II we present the
Jaynes-Cummings Hamiltonian describing the dynamics of two qubits
interacting with continuum of modes in a one- dimensional open
waveguide. We choose a generic state function on the Hilbert space
truncated to the single-excitation subspace. In Section III the
general time-dependent equations for qubits' amplitudes,
$\beta_1(t)$, $\beta_2(t)$, and photon forward, $\gamma(\omega,t)$
and backward $\delta(\omega,t)$ radiation amplitudes are derived
for the process when a single-photon pulse is scattered by a
system of two identical qubits. The main  results of the paper are
presented in Section IV where we calculated the spatio-temporal
structure of electric field for photon forward and backward
travelling waves and discuss the spectral features of the
transmitted and reflected waves. Summary of our work is presented
in conclusion (Section V). Some technical details of the
calculations are given in Appendices A and B.

\section{The model}
We consider two identical qubits in a one- dimensional infinite
waveguide with $\Omega$ and $d$ being the qubit frequency and the
distance between them, respectively. On $x$-axis the qubits are
located at the points $x_1=0$ and $x_2=d$, respectively.

In the continuum mode representation this system can be described
by a Jaynes-Cummings Hamiltonian which accounts for the
interaction between qubits and electromagnetic field
\cite{Blow1990,Dom2002} (from now on we use the units $\hbar = 1$
throughout the paper, therefore, all energies are expressed in
frequency units):
  \begin{equation}\label{1}
\begin{gathered}
  H = \frac{1}
{2}\sum\limits_{n = 1}^2 {} \left( {1 + \sigma _Z^{(n} } \right)\Omega  \hfill \\
   + \int\limits_0^\infty  {} \omega a^ +  (\omega )a(\omega )d\omega  + \int\limits_0^\infty  {} \omega b^ +  (\omega )b(\omega )d\omega  \hfill \\
   + \int\limits_0^\infty  {} g(\omega )a^ +  (\omega )\sigma _ - ^{(1)} d\omega  + \int\limits_0^\infty  {} g(\omega )e^{ - ikd} a^ +  (\omega )\sigma _ - ^{(2)} d\omega  \hfill \\
   + \int\limits_0^\infty  {} g(\omega )a(\omega )\sigma _ + ^{(1)} d\omega  + \int\limits_0^\infty  {} g(\omega )e^{ikd} a(\omega )\sigma _ + ^{(2)} d\omega  \hfill \\
   + \int\limits_0^\infty  {} g(\omega )b^ +  (\omega )\sigma _ - ^{(1)} d\omega  + \int\limits_0^\infty  {} g(\omega )e^{ikd} b^ +  (\omega )\sigma _ - ^{(2)} d\omega  \hfill \\
   + \int\limits_0^\infty  {} g(\omega )b(\omega )\sigma _ + ^{(1)} d\omega  + \int\limits_0^\infty  {} g(\omega )e^{ - ikd} b(\omega )\sigma _ + ^{(2)} d\omega  \hfill \\
\end{gathered}
\end{equation}

where $k=\omega/v_g$, $v_g$ is the group velocity of
electromagnetic waves, which in our calculations is taken to be
the vacuum light speed, $v_g=3\times 10^8$m/s; $\omega$ is a
photon frequency. $\sigma^{(n)}_z, n=1,2$ is a Pauli spin
operator, $\sigma_-^{(n)} = {\left|g\right\rangle_{nn}}
\left\langle e \right|$ and $\sigma _ + ^{(n)} = {\left| e
\right\rangle _{nn}} \left\langle g \right|$ are the lowering and
raising atomic operators which lower or raise a state of the $n$th
qubit. A spin operator $\sigma_z^{(n)}=|e\rangle_{nn}\langle
e|-|g\rangle_{nn}\langle g|$.

The photon creation and annihilation operators $a^\dag(\omega)$,
$a(\omega)$, and $b^\dag(\omega)$, $b(\omega)$ describe forward and
backward scattering waves, respectively. They are independent of
each other and satisfy the usual continuous-mode commutation
relations \cite{Blow1990}:
  \begin{equation}\label{comut}
\left[ {a(\omega ),a^\dag  (\omega ')} \right] = \left[ {b(\omega ),b^\dag (\omega ')} \right] = \delta(\omega - \omega ').
  \end{equation}

 The quantity $g(\omega)$ in
(\ref{1}) is the coupling between qubit and the photon field in a
waveguide \cite{Dom2002}:
  \begin{equation}\label{g}
g(\omega ) = \sqrt {\frac{{\omega p^2 }} {{4\pi \varepsilon _0
\hbar v_g S}}},
  \end{equation}
where $p$ is the off diagonal matrix element of a dipole operator,
$S$ is the effective transverse cross section of the modes in
one-dimensional waveguide. We assume that the coupling is the same
for forward and backward waves.

Note that the dimension of the coupling constant $g(\omega)$ is
not a frequency, $\omega$ but a square of frequency,
$\sqrt{\omega}$, and, as it follows from (\ref{comut}), the
dimension of creation and destruction operators is
$1/\sqrt{\omega}$.

Below we consider a single-excitation subspace with either a
single photon is in a waveguide and two qubits are in the ground
state, or there are no photons in a waveguide with the only one
(first or second) qubit in the chain being excited. Therefore, we
truncate Hilbert space to the following states:
  \begin{equation}\label{H-space}
\begin{array}{l}
 \left| {G,1_k } \right\rangle  = \left| {g_1 ,g_2 } \right\rangle  \otimes \left| {1_k } \right\rangle ;
 \\\\
 \left| {1,0_k } \right\rangle  = \left| {e_1, g_2 } \right\rangle  \otimes \left| {0_k } \right\rangle ;
\\\\
\left| {2,0_k } \right\rangle  = \left| {g_1, e_2} \right\rangle
\otimes \left| {0_k } \right\rangle ;
\end{array}
  \end{equation}

The Hamiltonian (\ref{1}) preserves the number of excitations
(number of excited qubits + number of photons). In our case the
number of excitations is equal to one. Therefore, at any time the
system will remain within a single-excitation subspace. It is
worth noting here that the Hilbert space is truncated with respect
to the number of photons, but the continuum modes are not
truncated as was done, for example, in \cite{Drob2000}.

The trial wave function of an arbitrary single-excitation state
can then be written in the form:
  \begin{equation}\label{3}
\begin{gathered}
  \Psi (t) = \sum\limits_{n = 1}^2 {} \beta_n (t)e^{-i\Omega t} \left| {n,0} \right\rangle  \hfill \\
   + \int\limits_0^\infty  {d\omega \gamma(\omega ,t) e^{-i\omega t} a^\dag (\omega )\left| {G,0} \right\rangle }  \hfill \\
   + \int\limits_0^\infty  {d\omega \delta (\omega ,t) e^{-i\omega t} b^\dag (\omega )\left| {G,0} \right\rangle },  \hfill \\
\end{gathered}
  \end{equation}
where $\beta_n(t)$  is the amplitude of $n$th qubit ($n=1,2$),
$\gamma(\omega,t)$ and $\delta(\omega,t)$ are single-photon
amplitudes which are related to a spectral density of forward and
backward radiation, respectively.

The function (\ref{3}) at any time is normalized to unity:

  \begin{equation}\label{4}
\sum\limits_{n = 1}^2 {\left| {\beta _n (t)} \right|^2 }  +
  \int\limits_0^\infty  {d\omega } \left| {\gamma (\omega ,t)}
  \right|^2  + \int\limits_0^\infty  {d\omega } \left| {\delta (\omega ,t)}
  \right|^2=1.
  \end{equation}

The initial conditions are as follows: all qubits are unexcited at
$t=0$, $\beta_n(0)=0$; reflected wave is absent,
$\delta(\omega,0)=0$; transmitted wave
$\gamma(\omega,0)\equiv\gamma_0(\omega)$, where $\gamma_0(\omega)$
is the incident pulse which for a single scattering photon is
assumed to be normalized to unity, $\int\limits_0^\infty  {d\omega
} \left| {\gamma _0 (\omega )} \right|^2=1$.

We note that as $t\rightarrow\infty$ the qubit amplitude
$\beta_n(t)\rightarrow 0$, therefore the equation (\ref{4})
reduces to:
  \begin{equation}\label{4a}
\int\limits_0^\infty  {d\omega } \left| {\gamma (\omega
,t\rightarrow\infty)}
  \right|^2  + \int\limits_0^\infty  {d\omega } \left| {\delta (\omega ,t\rightarrow\infty)}
  \right|^2=1.
  \end{equation}

\section{Equations for qubits' and photon amplitudes}
The Schrodinger equation $ i\frac{{d\Psi (t)}} {{dt}} = H\Psi (t)$
yields the equations for the amplitudes:
  \begin{equation}\label{5}
\begin{gathered}
  \frac{{d\beta _1 }}
{{dt}} =  - i\int\limits_0^\infty  {} d\omega g(\omega )\gamma (\omega ,t)e^{ - i(\omega  - \Omega )t}  \hfill \\
   - i\int\limits_0^\infty  {} d\omega g(\omega )\delta (\omega ,t)e^{ - i(\omega  - \Omega )t}  \hfill \\
\end{gathered}
\end{equation}

  \begin{equation}\label{6}
\begin{gathered}
  \frac{{d\beta _2 }}
{{dt}} =  - i\int\limits_0^\infty  {} d\omega g(\omega )\gamma (\omega ,t)e^{ikd} e^{ - i(\omega  - \Omega )t}  \hfill \\
   - i\int\limits_0^\infty  {} d\omega g(\omega )\delta (\omega ,t)e^{ - ikd} e^{ - i(\omega  - \Omega )t}  \hfill \\
\end{gathered}
\end{equation}

\begin{equation}\label{7}
\frac{{d\gamma (\omega ,t)}} {{dt}} =  - i\beta _1 (t)g(\omega
)e^{i(\omega  - \Omega )t}  - i\beta _2 (t)g(\omega )e^{i(\omega -
\Omega )t} e^{ - ikd}
\end{equation}
\begin{equation}\label{7a}
\frac{{d\delta (\omega ,t)}} {{dt}} =  - i\beta _1 (t)g(\omega
)e^{i(\omega  - \Omega )t}  - i\beta _2 (t)g(\omega )e^{i(\omega -
\Omega )t} e^{ikd}
\end{equation}

We assume the photon is incident from the left on the first qubit
at $t=0$, so that $\gamma(\omega,0)\equiv\gamma_0(\omega)$ and
$\delta(\omega,0)=0$. The qubits are assumed to be in the ground
state at $t=0$ so that  $\beta_1(0)=\beta_2(0)=0$.

Therefore, the formal solutions of equations (\ref{7}) and
(\ref{7a}) are as follow:

 \begin{equation}\label{8}
\begin{gathered}
  \gamma (\omega ,t) = \gamma _0 \left( \omega  \right) - ig(\omega )\int\limits_0^t {dt'} \beta _1 (t')e^{i\left( {\omega  - \Omega } \right)t'}  \hfill \\
   - ig(\omega )e^{ - ikd} \int\limits_0^t {dt'} \beta _2 (t')e^{i\left( {\omega  - \Omega } \right)t'}  \hfill \\
\end{gathered}
\end{equation}

 \begin{equation}\label{9}
\begin{gathered}
  \delta (\omega ,t) =  - ig(\omega )\int\limits_0^t {dt'} \beta _1 (t')e^{i\left( {\omega  - \Omega } \right)t'}  \hfill \\
   - ig(\omega )e^{ikd} \int\limits_0^t {dt'} \beta _2 (t')e^{i\left( {\omega  - \Omega } \right)t'}  \hfill \\
\end{gathered}
\end{equation}

Plugging equations (\ref{8}) and (\ref{9}) in (\ref{5}) and
(\ref{6}) we obtain the equations for qubits' amplitudes:

\begin{equation}\label{10}
\begin{gathered}
  \frac{{d\beta _1 }}
{{dt}} =  - i\int\limits_0^\infty  {d\omega } g(\omega )\gamma _0 \left( \omega  \right)e^{ - i(\omega  - \Omega )t}  \hfill \\
   - 2\int\limits_0^\infty  {d\omega } g^2 (\omega )\int\limits_0^t {dt'} \beta _1 (t')e^{ - i\left( {\omega  - \Omega } \right)(t - t')}  \hfill \\
   - 2\int\limits_0^\infty  {d\omega } g^2 (\omega )\cos kd\int\limits_0^t {dt'} \beta _2 (t')e^{ - i\left( {\omega  - \Omega } \right)(t - t')}  \hfill \\
\end{gathered}
\end{equation}

\begin{equation}\label{11}
\begin{gathered}
  \frac{{d\beta _2 }}
{{dt}} =  - i\int\limits_0^\infty  {d\omega } g(\omega )\gamma _0 \left( \omega  \right)e^{ikd} e^{ - i(\omega  - \Omega )t}  \hfill \\
   - 2\int\limits_0^\infty  {d\omega } g^2 (\omega )\cos kd\int\limits_0^t {dt'} \beta _1 (t')e^{ - i\left( {\omega  - \Omega } \right)(t - t')}  \hfill \\
   - 2\int\limits_0^\infty  {d\omega } g^2 (\omega )\int\limits_0^t {dt'} \beta _2 (t')e^{ - i\left( {\omega  - \Omega } \right)(t - t')}  \hfill \\
\end{gathered}
\end{equation}
In order to obtain analytical solutions for equations (\ref{10})
and (\ref{11}) we assume that the qubit amplitudes $\beta_1(t),
\beta_2(t)$ vary with a rate which is much less than the qubit
frequency $\Omega$. Therefore, the qubit amplitudes change little
in the time interval over which the remaining part of the
integrands have non-zero value $(t'\sim t)$. Therefore, we can
replace $\beta_1(t'), \beta_2(t')$  in the integrands by
$\beta_1(t), \beta_2(t)$  and take them out of the integrals. This
is called the Weisskopf-Wigner approximation, which is equivalent
to the Markov approximation: dynamics of $\beta_n(t)$ depends only
on time $t$ and not on $t'<t$, i.e., the system has no memory of
the past.

After some mathematical manipulations (details are given in
Appendix A) we arrive at the following equations for qubits'
amplitudes:
\begin{equation}\label{12}
\frac{{d\beta _1 }} {{dt}} =  - i\int\limits_0^\infty  {d\omega }
g(\omega )\gamma _0 \left( \omega  \right)e^{ - i(\omega  - \Omega
)t}  - \frac{\Gamma } {2}\beta _1 (t) - \frac{\Gamma }
{2}e^{ik_\Omega  d} \beta _2 (t)
\end{equation}

\begin{equation}\label{13}
\begin{gathered}
  \frac{{d\beta _2 }}
{{dt}} =  - i\int\limits_0^\infty  {d\omega } g(\omega )\gamma _0 \left( \omega  \right)e^{ik_\omega  d} e^{ - i(\omega  - \Omega )t}  \hfill \\
   - \frac{\Gamma }
{2}\beta _2 (t) - \frac{\Gamma }
{2}e^{ik_\Omega  d} \beta _1 (t) \hfill \\
\end{gathered}
\end{equation}
where $\Gamma=4\pi g^2(\Omega)$ is the rate of spontaneous
emission into a waveguide, $k_{\omega}=\omega/v_g$,
$k_{\Omega}=\Omega/v_g$.

It is not difficult to find from (\ref{12}), (\ref{13}) analytical
solutions for $\beta_n(t)$ with initial conditions $\beta_n(0)=0$:

\begin{equation}\label{14}
\beta _1 (t) = \frac{1} {2}\int\limits_0^t {d\tau } F_ +  (\tau
)e^{ - \Gamma _ +  (t - \tau )}  + \frac{1} {2}\int\limits_0^t
{d\tau } F_ -  (\tau )e^{ - \Gamma _ -  (t - \tau )}
\end{equation}

\begin{equation}\label{15}
\beta _2 (t) = \frac{1} {2}\int\limits_0^t {d\tau } F_ +  (\tau
)e^{ - \Gamma _ +  (t - \tau )}  - \frac{1} {2}\int\limits_0^t
{d\tau } F_ -  (\tau )e^{ - \Gamma _ -  (t - \tau )}
\end{equation}
where
\begin{equation}\label{16}
F_ \pm  (t) =  - i\int\limits_0^\infty  {d\omega } g(\omega
)\gamma _0 \left( \omega  \right)e^{ - i(\omega  - \Omega )t}
\left( {1 \pm e^{ik_\omega  d} } \right)
\end{equation}

\begin{equation}\label{17}
\Gamma _ \pm   = \frac{\Gamma } {2}\left( {1 \pm e^{ik_\Omega  d}
} \right)
\end{equation}

Below for the calculations we use the incident Gaussian pulse:

\begin{equation}\label{Gauss}
\gamma _0 (\omega ) = \left( {\frac{2} {{\pi \Delta ^2 }}}
\right)^{1/4} e^{ - \frac{{(\omega  - \omega _S )^2 }} {{\Delta ^2
}}}
\end{equation}
where $\Delta$ is the width of the pulse in the frequency domain.

We assume the Gaussian pulse (\ref{Gauss}) is sufficiently narrow
 ($\Delta\ll\omega_S$)  so that it can be approximated as a delta
pulse \cite{Green2023}:

\begin{equation}\label{Gauss1}
\gamma _0 (\omega ) =A \delta (\omega  - \omega _S )
\end{equation}
where $A= \left( {2\pi } \right)^{1/4} \sqrt \Delta$ is the
amplitude of incoming wave and $\omega_S$ is the frequency of the
incident photon.

For delta pulse (\ref{Gauss1}) we obtain final expressions for
qubits' amplitudes:

\begin{equation}\label{18}
\begin{gathered}
  \beta _1 (t) = \frac{1}{2}C_ +  \left( {e^{ - \Gamma _ +  t}  - e^{i\left( {\Omega  - \omega _s } \right)t} } \right) \hfill \\
   + \frac{1}
{2}C_ -  \left( {e^{ - \Gamma _ -  t}  - e^{i\left( {\Omega  - \omega _s } \right)t} } \right) \hfill \\
\end{gathered}
\end{equation}

\begin{equation}\label{19}
\begin{gathered}
  \beta _2 (t) = \frac{1}
{2}C_ +  \left( {e^{ - \Gamma _ +  t}  - e^{i\left( {\Omega  - \omega _s } \right)t} } \right) \hfill \\
   - \frac{1}
{2}C_ -  \left( {e^{ - \Gamma _ -  t}  - e^{i\left( {\Omega  - \omega _s } \right)t} } \right) \hfill \\
\end{gathered}
\end{equation}

where
\begin{equation}\label{20}
C_ \pm   = \frac{{A \sqrt {\frac{\Gamma } {{4\pi }}} \left( {1 \pm
e^{ik_{\omega _S } d} } \right)}} {{\left( {\Omega  - \omega _s }
\right) - i\Gamma _ \pm }}
\end{equation}

The photon radiation amplitudes $\gamma(\omega,t))$,
$\delta(\omega,t)$ are obtained by plugging the expressions
(\ref{18}), (\ref{19}) for $\beta_1(t)$, $\beta_2(t)$ directly in
(\ref{8}) and (\ref{9}). The calculations of corresponding
integrals in (\ref{8}) and (\ref{9}) then yield the following
result:
\begin{equation}\label{gamma}
    \gamma(\omega,t)=\gamma_0(\omega)+\gamma_1(\omega,t)
\end{equation}
where
\begin{equation}\label{21}
\begin{gathered}
  \gamma_1 (\omega ,t) =  - \frac{1}
{2}\sqrt {\frac{\Gamma }
{{4\pi }}} \left( {1 + e^{ - ik_\omega  d} } \right)C_ +  D_ +   \hfill \\
   - \frac{1}
{2}\sqrt {\frac{\Gamma }
{{4\pi }}} \left( {1 - e^{ - ik_\omega  d} } \right)C_ -  D_ -   \hfill \\
\end{gathered}
\end{equation}

\begin{equation}\label{22}
\begin{gathered}
  \delta (\omega ,t) =  - \frac{1}
{2}\sqrt {\frac{\Gamma }
{{4\pi }}} \left( {1 + e^{ik_\omega  d} } \right)C_ +  D_ +   \hfill \\
   - \frac{1}
{2}\sqrt {\frac{\Gamma }
{{4\pi }}} \left( {1 - e^{ik_\omega  d} } \right)C_ -  D_ -   \hfill \\
\end{gathered}
\end{equation}
where
\begin{equation}\label{23}
D_ \pm   = \frac{{e^{i(\omega  - \Omega  + i\Gamma _ \pm  )t}  -
1}} {{\omega  - \Omega  + i\Gamma _ \pm  }} - \frac{{e^{i(\omega -
\omega _S )t}  - 1}} {{\omega  - \omega _S }}
\end{equation}

\section{Space-time structure of the scattered field}
\subsection{Forward scattering field}
The photon wave packet for forward propagating field is given by

\begin{equation}\label{Fw}
 u(x,t) = \int\limits_0^\infty  {} d\omega \gamma (\omega ,t)e^{ i\frac{\omega }
{{v_g }}(x - v_g t)}=u_0(x,t)+u_1(x,t)
\end{equation}

where
\begin{equation}\label{Fw1}
 u_0(x,t)
   = \int\limits_0^\infty  {} d\omega \gamma _0 (\omega )e^{  i\frac{\omega }
{{v_g }}(x - v_g t)}
\end{equation}

\begin{equation}\label{Fw2}
u_1(x,t) = \int\limits_0^\infty  {} d\omega \gamma _1 (\omega
,t)e^{  i\frac{\omega } {{v_g }}(x - v_g t)}
\end{equation}

where $\gamma_0(\omega)$ is given in (\ref{Gauss}).

In equations (\ref{Fw1}) $x$ takes any value, both positive and
negative, while in equation (\ref{Fw2}) $x>0$. In both equations
$x-v_gt<0$. This condition insures the causality of the forward
propagating field which appears at the point $x$ not until the
signal travels the distance $x$ after the scattering. Therefore,
the expression (\ref{Fw2}) describes the field both between the
qubits, $0<x<d$ and behind the second qubit, $x>d$.

The quantity $u_0(x,t)$ in (\ref{Fw1}) reads:
\begin{equation}\label{sum1}
u_0(x,t)  = A e^{ i\frac{{\omega _S }} {{v_g }}(x - v_g t)}
\end{equation}
where we use a small $\Delta$ approximation (\ref{Gauss1}).

From (\ref{21}) the quantity $u_1(x,t)$, (\ref{Fw2}) can be
written as follows:
\begin{equation}\label{35}
\begin{gathered}
  u_1 (x,t) = \hfill\\ - \frac{1}
{2}\sqrt {\frac{\Gamma }
{{4\pi }}} C_ +  \left( {I_1^ +  (x,t) + I_1^ +  (x - d,t) - I_2 (x,t) - I_2 (x - d,t)} \right) \hfill \\
   - \frac{1}
{2}\sqrt {\frac{\Gamma }
{{4\pi }}} C_ -  \left( {I_1^ -  (x,t) - I_1^ -  (x - d,t) - I_2 (x,t) + I_2 (x - d,t)} \right) \hfill \\
\end{gathered}
\end{equation}
where

\begin{equation}\label{36}
 I_1^ \pm  (x,t) = \int\limits_0^\infty  {d\omega }
\frac{{e^{i(\omega  - \Omega  + i\Gamma _ \pm  )t}  - 1}} {{\omega
- \Omega  + i\Gamma _ \pm  }}e^{i\frac{\omega } {{v_g }}\left( {x
- v_g t} \right)}
\end{equation}

\begin{equation}\label{37}
 I_1^ \pm  (x - d,t) = \int\limits_0^\infty  {d\omega }
\frac{{e^{i(\omega  - \Omega  + i\Gamma _ \pm  )t}  - 1}} {{\omega
- \Omega  + i\Gamma _ \pm  }}e^{i\frac{\omega } {{v_g }}\left( {x
- d - v_g t} \right)}
\end{equation}
where $\Gamma_{\pm}$ are given in (\ref{17}).

\begin{equation}\label{38}
 I_2 (x,t) = \int\limits_0^\infty  {d\omega } \frac{{e^{i(\omega -
\omega _S )t}  - 1}} {{\omega  - \omega _S }}e^{i\frac{\omega }
{{v_g }}\left( {x - v_g t} \right)}
\end{equation}

\begin{equation}\label{39}
I_2 (x - d,t) = \int\limits_0^\infty  {d\omega }
\frac{{e^{i(\omega  - \omega _S )t}  - 1}} {{\omega  - \omega _S
}}e^{i\frac{\omega } {{v_g }}\left( {x - d - v_g t} \right)}
\end{equation}

\subsubsection{Forward scattering field behind second qubit}

First we consider the forward scattering field behind the second
qubit, $x>d$ if $k_{\Omega}d\neq n\pi$, where $n$ is integer. For
this case, the integrals (\ref{36}), (\ref{38}) have been
calculated in \cite{Green2023}:

\begin{equation}\label{40}
\begin{gathered}
 I_1^ \pm  (x,t) = e^{ - i\Omega _ \pm  \;t} e^{i\frac{x} {{v_g
}}\Omega _ \pm  } E_1\left( {\frac{x} {{v_g }}\Omega _ \pm  }
\right) + 2\pi ie^{i\frac{{\Omega _ \pm  }} {{v_g }}(x - v_g
t)}\hfill\\
 - e^{ - i\frac{{\left| {x - v_g t} \right|}} {{v_g
}}\Omega _ \pm } E_1\left( { - i\frac{{\left| {x - v_g t}
\right|}} {{v_g }}\Omega _ \pm  } \right)\;
\end{gathered}
\end{equation}

\begin{equation}\label{42}
 \begin{gathered}
  I_2 (x,t) = e^{i\frac{{\omega _s }}
{{v_g }}\left( {x - v_g t} \right)} \left\{ {2\pi i -
\,\rm{ci}\left( {\omega _s \frac{x} {{v_g }}} \right) +
\emph{i}\,\rm{si}\left( {\omega _s \frac{x}
{{v_g }}} \right)} \right. \hfill \\
  \left. { + \,\rm{ci}\left( {\omega _s \left| {\frac{{x - v_g t}}
{{v_g }}} \right|} \right) + \emph{i}\,\rm{si}\left( {\omega _s
\left| {\frac{{x - v_g t}}
{{v_g }}} \right|} \right)} \right\} \hfill \\
\end{gathered}
\end{equation}

The integrals $I_1^{\pm}(x-d,t)$ and $I_2(x-d,t)$ are obtained by
replacing $x$ with $x-d$ in the expressions for $I_1^{\pm}(x,t)$
and $I_2(x,t)$.

\begin{equation}\label{41}
\begin{gathered}
I_1^ \pm  (x - d,t) = e^{ - i\Omega _ \pm  \;t} e^{i\frac{{x - d}}
{{v_g }}\Omega _ \pm  } E_1\left( {\frac{{x - d}} {{v_g }}\Omega _
\pm  } \right)\hfill\\ + 2\pi ie^{i\frac{{\Omega _ \pm  }} {{v_g
}}(x - d - v_g t)}\hfill\\
 - e^{ - i\frac{{\left| {x - d - v_g t}
\right|}} {{v_g }}\Omega _ \pm  } E_1\left( { - i\frac{{\left| {x
- d - v_g t} \right|}} {{v_g }}\Omega _ \pm  } \right)\;
\end{gathered}
\end{equation}

\begin{equation}\label{43}
\begin{gathered}
\begin{gathered}
  I_2 (x - d,t) = e^{i\frac{{\omega _s }}
{{v_g }}\left( {x - d - v_g t} \right)} 2\pi i \hfill \\
   + e^{i\frac{{\omega _s }}
{{v_g }}\left( {x - d - v_g t} \right)} \left\{ { -
\,\rm{ci}\left( {\omega _s  {\frac{{x - d}} {{v_g }}} } \right) +
\emph{i}\,\rm{si}\left( {\omega _s \frac{{x - d}}
{{v_g }}} \right)} \right. \hfill \\
  \left. { + \,\rm{ci}\left( {\omega _s \left| {\frac{{x - d - v_g t}}
{{v_g }}} \right|} \right) + \emph{i}\,\rm{si}\left( {\omega _s
\left| {\frac{{x - d - v_g t}}
{{v_g }}} \right|} \right)} \right\} \hfill \\
\end{gathered}
\end{gathered}
\end{equation}

where $\Omega_{\pm}=\Omega-i\Gamma_{\pm}$, $x>d$, $x-v_gt<0$,
$x-d-v_gt<0$.

In expressions (\ref{40})-(\ref{43}) the quantities $E_1(z)$,
$\rm{ci}(z)$, and $\rm{si}(z)$ are the exponential integral
function, cosine, and sine integrals, respectively, the properties
of which are described in Appendix B.

 The integrals $I_1^{\pm}(x,t)$, $I_1^{\pm}(x-d,t)$ are the damping parts
 of the scattered field which represents spontaneous emission of the excited
qubits with the rates $\Gamma_{\pm}$, which describe the
exponential decay of the corresponding collective states of a
two-qubt system. Depending on the value $k_{\Omega}d$, the damping
rates $\Gamma_{\pm}$ may change from superradiance
($\Gamma_{\pm}\approx\Gamma$) to subradiance
($\Gamma_{\mp}\ll\Gamma$) emission. The case for which one of the
rates $\Gamma_{\pm}$ equals zero will be considered later.

The integrals $I_2(x,t)$, $I_2(x-d,t)$ consist of two parts: the
time-dependent part which decays as the inverse powers of $t$, and
a lossless steady state solution of the scattered field which
survives as the time after the scattering tends to infinity. As
both the distance from the qubit and the time after the scattering
tend to infinity, these scattering fields die out, leaving only
the plane-wave stationary solution.

We investigate equation (\ref{35}) when $t\rightarrow\infty$. In
this case, the quantities $I_1^{\pm}(x,t), I_1^{\pm}(x-d,t)$ tend
to zero exponentially, while in the quantities $I_2(x,t),
I_2(x-d,t)$ the time-dependent terms tend to zero as the inverse
powers of $t$. Therefore, for $t\rightarrow\infty$ we obtain from
(\ref{35}):
\begin{widetext}
\begin{equation}\label{44}
\begin{gathered}
\begin{gathered}
  u(x,t \to \infty ) = e^{i\frac{{\omega _S }}
{{v_g }}\left( {x - v_g t} \right)} \left[ {A  + \frac{1} {2}\sqrt
{\frac{\Gamma } {{4\pi }}} (C_ +   + C_ -  )\left( {2\pi i -
\,\rm{ci}\left( {\omega _s \frac{\emph{x}} {{v_g }}} \right) +
\emph{i}\,\rm{si}\left( {\omega _s \frac{\emph{x}}
{{v_g }}} \right)} \right)} \right] \hfill \\
   + e^{i\frac{{\omega _S }}
{{v_g }}\left( {x - v_g t} \right)} \left[ {\frac{1} {2}\sqrt
{\frac{\Gamma } {{4\pi }}} e^{ - i\frac{{\omega _s }} {{v_g }}d}
(C_ +   - C_ -  )\left( {2\pi i - \,\rm{ci}\left( {\omega _s
{\frac{{\emph{x} - \emph{d}}} {{v_g }}} } \right) +
\emph{i}\,\rm{si}\left( {\omega _s \frac{{\emph{x} - \emph{d}}}
{{v_g }}} \right)} \right)} \right] \hfill \\
\end{gathered}
\end{gathered}
\end{equation}
where $x>d$, $k_{\Omega}d\neq n\pi$.
\end{widetext}

As a general note to equation (\ref{44}), we observe that the
field behind the second qubit is not zero for finite $x$. As $x$
is increased the $x$-dependent part of (\ref{44}) tends to zero as
the inverse power of $x$ (see (\ref{Asymp1}) in Appendix B). In
principle, the steady state field energy
$|u(x,t\rightarrow\infty|^2$ exists in all points of
one-dimensional space.

When calculating (\ref{44}) some caution should be paid if $x$ is
very near to the qubit location, where cosine integral diverges.
This nonphysical effect arises from the point-like model of a
qubit adopted in this paper. In order to avoid this effect in all
calculations where we study the $x$-dependence of the field, we
start the numerics at the distance $\Delta x=0.05d$ from a qubit
where the influence of cosine integral becomes negligible.

\begin{figure}
  \includegraphics[width=8 cm]{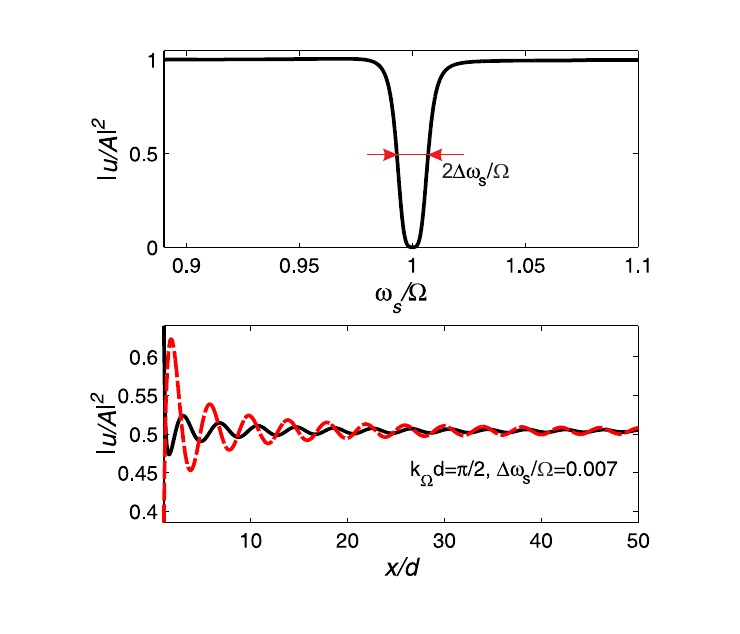}\\
  \caption{Transmitted energy behind the second qubit. Upper panel
  shows the resonance curve calculated from (\ref{44}) for
  $k_{\Omega}d=\pi/2$ ($d=0.015$m)
 at $x=3d$ (at the distance $2d$ from the second qubit).
  Spatial dependence of  transmitted energy (\ref{44}) is shown in the lower panel
  for off-resonant conditions, $(\omega_S-\Omega)/\Omega=0.007$, (solid black  line)
   and $(\omega_S-\Omega)/\Omega=-0.007$ (dashed red line).
  $\Gamma/\Omega=0.01$, $\Omega/2\pi=5$ GHz, $\lambda=6$
cm. A starting point for the calculation of spatial dependence is
taken at $x=1.05d$.}\label{Fig6}
\end{figure}

As it follows from (\ref{44}) spatial effects can persist on the
scale of several wavelengthes. We show in Fig.\ref{Fig6} the
variation of transmitted energy in $x$ if the frequency is detuned
from the resonance by a half width.

For finite $x$ the transmitted energy at resonance
($\omega_S=\Omega$) can be obtained directly from (\ref{44}).
\begin{equation}\label{45a}
\frac{{\left| {u(x,t \to \infty )} \right|^2 }} {{A^2 }} =
\frac{1} {{4\pi ^2 }}\,\left( {\rm{ci}^2 \left( {\Omega \frac{x}
{{v_g }}} \right) + \rm{si}^2 \left( {\Omega \frac{x} {{v_g }}}
\right)} \right)
\end{equation}
where $x>d$, $k_{\Omega}d\neq n\pi$.

If in (\ref{44}) $x\rightarrow\infty$ we obtain the transmittance:
\begin{equation}\label{45}
\begin{gathered}
  \frac{{\left| {u(x \to \infty ,t \to \infty )} \right|^2 }}
{{A^2 }} = \left| 1 \right. \hfill \\
  \left. { + \frac{1}
{{2A }}\sqrt {\frac{\Gamma } {{4\pi }}} 2\pi i\left( {C_ +  (1 +
e^{ - i\frac{{\omega _s }} {{v_g }}d} ) + C_ -  (1 - e^{ -
i\frac{{\omega _s }}
{{v_g }}d} )} \right)} \right|^2  \hfill \\
\end{gathered}
\end{equation}
where $A=(2\pi)^{1/4}\sqrt{\Delta}$ is the amplitude of the
incident pulse.

Even though (\ref{45}) is derived from (\ref{44}) for
$k_{\Omega}d\neq n\pi$, we will show below that (\ref{45}) is
valid for any value of $k_{\Omega}d$.

Using the explicit expressions for $C_{\pm}$ (\ref{20}) it is not
difficult to show from (\ref{45}) that in resonance,
$\omega_S=\Omega$, far from second qubit the transmitted field is
absent, $u(x\rightarrow\infty,t\rightarrow\infty)=0$, which is in
accordance with the result of quasi stationary theory.

We may compare (\ref{45}) with the non-Markovian transmission
which accounts for retardation effects \cite{Green2015, Liao2015}:

\begin{equation}\label{TR}
|T|^2  = \left| {\frac{{\left( {\omega  - \Omega } \right)^2 }}
{{\left( {\omega  - \Omega  + i\frac{\Gamma } {2}} \right)^2  +
\frac{{\Gamma ^2 }} {4}e^{2ik_\omega  d} }}} \right|^2
\end{equation}

\begin{figure}
  \includegraphics[width=8 cm]{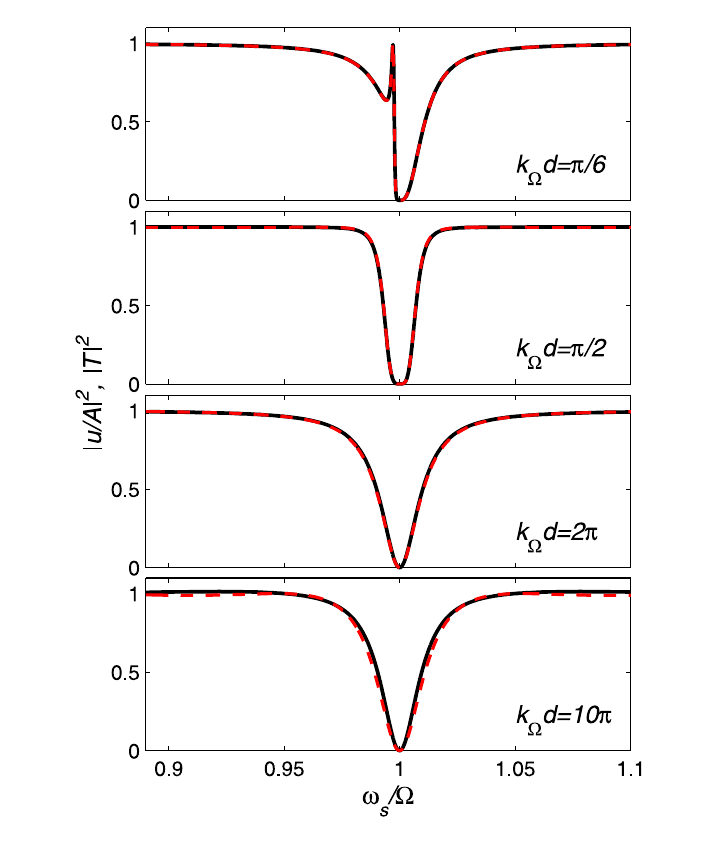}\\
  \caption{The comparison of the Markovian transmittance
  (\ref{45}), (black solid line)
  with non-Markovian expression (\ref{TR}) (dashed red line).
  $\Gamma/\Omega=0.01$, $\Omega/2\pi=5$GHz.}\label{Fig2}
\end{figure}

\begin{figure}
  \includegraphics[width=8 cm]{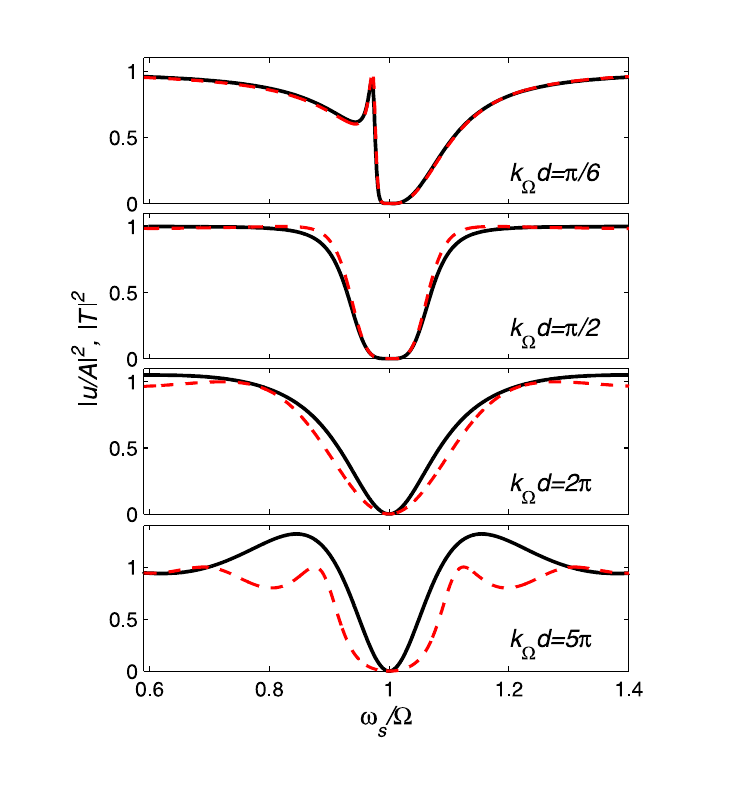}\\
  \caption{The comparison of the Markovian transmittance
  (\ref{45}), (black solid line)
  with non-Markovian expression (\ref{TR}) (dashed red line).
  $\Gamma/\Omega=0.1$, $\Omega/2\pi=5$GHz.}\label{Fig3}
\end{figure}

The results of the comparison are shown in Fig.\ref{Fig2} and
Fig.\ref{Fig3}. For $\Gamma/\Omega=0.01$ the frequency dependence
(\ref{45}) practically coincides with that of (\ref{TR}) up to
 $k_{\Omega}d\approx 10\pi$. However, as the value of
$\Gamma$ increases, the departure between two transmittances
increases more rapidly and a significant difference is seen for
$k_{\Omega}d=5\pi$ (see Fig.\ref{Fig3}).

\subsubsection{Forward scattering field behind the second qubit for
$k_{\Omega}d=n\pi$}

In this case, one of the spontaneous rate emissions,
$\Gamma_{\pm}$ is zero. For $k_{\Omega}d=n\pi$ where $n$ is even
$\Gamma_+=\Gamma, \Gamma_-=0$. For $k_{\Omega}d=n\pi$ where $n$ is
odd $\Gamma_+=0, \Gamma_-=\Gamma$. Below for definiteness we take
$k_{\Omega}d=n\pi$ where $n$ is even number. Then, the photon
radiation amplitudes are given by the expressions (\ref{21}),
(\ref{22}) where

\begin{equation}\label{46}
D_ +   = \frac{{e^{i(\omega  - \Omega  + i\Gamma   )t}  - 1}}
{{\omega  - \Omega  + i\Gamma  }} - \frac{{e^{i(\omega - \omega _S
)t}  - 1}} {{\omega  - \omega _S }}
\end{equation}

\begin{equation}\label{47}
D_ -   = \frac{{e^{i(\omega  - \Omega)  t}  - 1}} {{\omega  -
\Omega   }} - \frac{{e^{i(\omega - \omega _S )t} - 1}} {{\omega -
\omega _S }}
\end{equation}

\begin{equation}\label{48}
C_ +   = \frac{{A\sqrt {\frac{\Gamma } {{4\pi }}} \left( {1 +
e^{ik_{\omega _S } d} } \right)}} {{\Omega  - \omega _s  - i\Gamma
}},\;C_ -   = \frac{{A\sqrt {\frac{\Gamma } {{4\pi }}} \left( {1 -
e^{ik_{\omega _S } d} } \right)}} {{\Omega  - \omega _s }}
\end{equation}

In this case, the quantity $u_1(x,t)$, (\ref{Fw2}) can be written
as follows:
\begin{equation}\label{49}
\begin{gathered}
  u_1 (x,t) = \hfill\\ - \frac{1}
{2}\sqrt {\frac{\Gamma }
{{4\pi }}} C_ +  \left( {I_1  (x,t) + I_1  (x - d,t) - I_2 (x,t) - I_2 (x - d,t)} \right) \hfill \\
   - \frac{1}
{2}\sqrt {\frac{\Gamma }
{{4\pi }}} C_ -  \left( {I_3  (x,t) - I_3  (x - d,t) - I_2 (x,t) + I_2 (x - d,t)} \right) \hfill \\
\end{gathered}
\end{equation}
where $C_{\pm}$ are given in (\ref{48}) and

\begin{equation}\label{50}
 I_1  (x,t) = \int\limits_0^\infty  {d\omega }
\frac{{e^{i(\omega  - \Omega  + i\Gamma  )t}  - 1}} {{\omega -
\Omega  + i\Gamma   }}e^{i\frac{\omega } {{v_g }}\left( {x - v_g
t} \right)}
\end{equation}

\begin{equation}\label{51}
 I_1  (x - d,t) = \int\limits_0^\infty  {d\omega }
\frac{{e^{i(\omega  - \Omega  + i\Gamma   )t}  - 1}} {{\omega -
\Omega  + i\Gamma  }}e^{i\frac{\omega } {{v_g }}\left( {x - d -
v_g t} \right)}
\end{equation}

\begin{equation}\label{52}
 I_3 (x,t) = \int\limits_0^\infty  {d\omega } \frac{{e^{i(\omega -
\Omega )t}  - 1}} {{\omega  - \Omega }}e^{i\frac{\omega } {{v_g
}}\left( {x - v_g t} \right)}
\end{equation}

\begin{equation}\label{53}
I_3 (x - d,t) = \int\limits_0^\infty  {d\omega }
\frac{{e^{i(\omega  - \Omega  )t}  - 1}} {{\omega  - \Omega
}}e^{i\frac{\omega } {{v_g }}\left( {x - d - v_g t} \right)}
\end{equation}

The quantities $I_1(x,t), I_1(x-d,t)$ in (\ref{49}) are given by
the expressions $I_1^+(x,t)$, (\ref{40}) and $I_1^+(x-d,t)$,
(\ref{41}) respectively, where $\Omega_+$ is replaced by
$\Omega-i\Gamma$. The quantities $I_2(x,t), I_2(x-d,t)$ are given
in (\ref{42}) , (\ref{43}), and the quantities $I_3(x,t),
I_3(x-d,t)$ are given by the expressions (\ref{42}) , (\ref{43}),
respectively, where $\omega_S$ is replaced by $\Omega$.

We investigate the equation (\ref{49}) when $t\rightarrow\infty$.
In this case, the quantities $I_1(x,t), I_1(x-d,t)$ tend to zero
exponentially and in the quantities $I_2(x,t), I_2(x-d,t)$,
$I_3(x,t), I_3(x-d,t)$ the time-dependent terms also tend to zero
as the inverse powers of $t$. Therefore, for $t\rightarrow\infty$
we obtain from (\ref{49}) the steady state expression:

\begin{widetext}
\begin{equation}\label{54}
\begin{gathered}
  u(x,t \to \infty ) = e^{i\frac{{\omega _S }}
{{v_g }}\left( {x - v_g t} \right)} \left[ {A + \frac{1} {2}\sqrt
{\frac{\Gamma } {{4\pi }}} (C_ +   + C_ -  )\left( {2\pi i -
\,\rm{ci}\left( {\omega _s \frac{\emph{x}} {{v_g }}} \right) +
\emph{i}\,\rm{si}\left( {\omega _s \frac{\emph{x}}
{{v_g }}} \right)} \right)} \right] \hfill \\
   + e^{i\frac{{\omega _S }}
{{v_g }}\left( {x - v_g t} \right)} \left[ {\frac{1} {2}\sqrt
{\frac{\Gamma } {{4\pi }}} e^{ - i\frac{{\omega _s }} {{v_g }}d}
(C_ +   - C_ -  )\left( {2\pi i - \,\rm{ci}\left( {\omega _s
 {\frac{{\emph{x - d}}} {{v_g }}}} \right) +
\emph{i}\,\rm{si}\left( {\omega _s \frac{{\emph{x - d}}}
{{v_g }}} \right)} \right)} \right] \hfill \\
   + e^{i\frac{\Omega }
{{v_g }}\left( {x - v_g t} \right)} \frac{1} {2}\sqrt
{\frac{\Gamma } {{4\pi }}} C_ -  \left[ {\,\rm{ci}\left( {\Omega
\frac{\emph{x}} {{v_g }}} \right) - \emph{i}\,\rm{si}\left(
{\Omega \frac{\emph{x}} {{v_g }}} \right) + \left( { -
\,\rm{ci}\left( {\Omega
 {\frac{{\emph{x - d}}} {{v_g }}} } \right) +
\emph{i}\,\rm{si}\left( {\Omega \frac{{\emph{x - d}}}
{{v_g }}} \right)} \right)} \right] \hfill \\
\end{gathered}
\end{equation}
where $x>d$. From (\ref{54}) we obtain for $\omega_S=\Omega$ the
equation which describes the dependence of the peak value of the
transmitted resonace line on the distance $x-d$ from the second
qubit.

\begin{equation}\label{54a}
\frac{{\left| {u(x,t \to \infty )} \right|^2 }} {{A^2 }} =
\frac{1} {{4\pi ^2 }}\left[ {\left( {\rm{ci}\left( {2\pi
\frac{{\left| \emph{x} \right|}} {\emph{d}}} \right) + ci\left(
{2\pi \frac{{\left| {\emph{x} - \emph{d}} \right|}} {\emph{d}}}
\right)} \right)^2  + \left( {\rm{si}\left( {2\pi \frac{{\left|
\emph{x} \right|}} {\emph{d}}} \right) + si\left( {2\pi
\frac{{\left| {\emph{x} - \emph{d}} \right|}} {\emph{d}}} \right)}
\right)^2 } \right]
\end{equation}
where $x>d$, $k_{\Omega}d=2\pi$.
\end{widetext}

If  $x\rightarrow\infty$, all sine and cosine integrals in
(\ref{54}) tend to zero and we obtain for the transmittance the
equation (\ref{45}) where $C_{\pm}$ are given in (\ref{48}).
Therefore, the transmittance (\ref{45}) is valid for any
$k_{\Omega}d$.

\begin{figure}
  \includegraphics[width=8 cm]{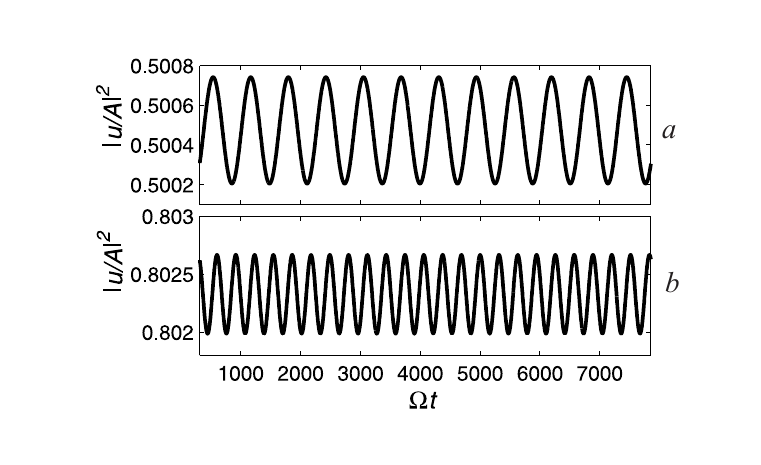}\\
  \caption{Beatings of the field energy behind the second qubit.
  Upper panel, $(\omega_S-\Omega)/\Omega=0.01$. Lower panel $(\omega_S-\Omega)/\Omega=0.02$.
  $x=2d$, $\Omega/2\pi=5$GHz, $\Gamma/\Omega=0.01$, $k_{\Omega}d=2\pi$, $d=0.06$m.}\label{Fig7}
\end{figure}

As is seen from (\ref{54}), at some specific point $x_0$ the field
energy exhibits the beatings with the detuning frequency
$\omega_S-\Omega$,
$|u(x_0,t\rightarrow\infty|^2\approx\cos({\omega_S-\Omega})t$, and
period $T=2\pi/(\omega_s-\Omega)$. The amplitude of these beatings
dies out as $x\rightarrow\infty$. The beatings for two detunings
are shown in Fig.\ref{Fig7} at the distance $d$ from the second
qubit.

It is not difficult to perform similar calculations for
$k_{\Omega}d=n\pi$ if $n$ is odd number. For this case we obtain
for $u_1(x,t)$ the expression which is similar to (\ref{49}).

\begin{equation}\label{55}
    \begin{gathered}
  u_1 (x,t) = \hfill\\ - \frac{1}
{2}\sqrt {\frac{\Gamma }
{{4\pi }}} C_ +  \left( {I_3  (x,t) + I_3  (x - d,t) - I_2 (x,t) - I_2 (x - d,t)} \right) \hfill \\
   - \frac{1}
{2}\sqrt {\frac{\Gamma }
{{4\pi }}} C_ -  \left( {I_1  (x,t) - I_1  (x - d,t) - I_2 (x,t) + I_2 (x - d,t)} \right) \hfill \\
\end{gathered}
\end{equation}

where
\begin{equation}\label{56}
C_ +   = \frac{{A\sqrt {\frac{\Gamma } {{4\pi }}} \left( {1 +
e^{ik_{\omega _S } d} } \right)}} {{\Omega  - \omega _s }},\;C_ -
= \frac{{A\sqrt {\frac{\Gamma } {{4\pi }}} \left( {1 -
e^{ik_{\omega _S } d} } \right)}} {{\Omega  - \omega _s-i\Gamma }}
\end{equation}
In this case we obtain for $u(x,t\rightarrow\infty)$ the
expression which is analogous to equation (\ref{54}) with the only
exception: in the third line in (\ref{54}) $C_-$ should be
replaced with $C_+$. Here, in the limit $x\rightarrow\infty$ we
obtain for the transmittance the equation (\ref{45}) where
$C_{\pm}$ are given in (\ref{56}).

In general, we may conclude that the transmittance (\ref{45}) is
valid for any value of $k_{\Omega}d/\pi$, integer or not integer.
However, as is seen from (\ref{44}) and (\ref{54}), the
expressions for the fields for these two cases are different.

\subsection{Backward scattering field}
The photon wave packet for backward propagating field is given by

\begin{equation}\label{Bw}
 v(x,t) = \int\limits_0^\infty  {} d\omega \delta (\omega ,t)e^{- i\frac{\omega }
{{v_g }}(x + v_g t)}
\end{equation}
     where $x<0$ and $x+v_g t>0$.

This quantity can be written similar to equation (\ref{35}).
\begin{widetext}
\begin{equation}\label{57}
\begin{gathered}
  {v} (x,t) =
   - \frac{1}
{2}\sqrt {\frac{\Gamma }
{{4\pi }}} C_ +  \left( {J_1^ +  (x,t) + J_1^ +  (x - d,t) - J_2 (x,t) - J_2 (x - d,t)} \right) \hfill \\
   - \frac{1}
{2}\sqrt {\frac{\Gamma }
{{4\pi }}} C_ -  \left( {J_1^ -  (x,t) - J_1^ -  (x - d,t) - J_2 (x,t) + J_2 (x - d,t)} \right) \hfill \\
\end{gathered}
\end{equation}
\end{widetext}
where
\begin{equation}\label{58}
J_1^ \pm  (x,t) = \int\limits_0^\infty  {d\omega }
\frac{{e^{i(\omega  - \Omega  + i\Gamma _ \pm  )t}  - 1}} {{\omega
- \Omega  + i\Gamma _ \pm  }}e^{ - i\frac{\omega } {{v_g }}\left(
{x + v_g t} \right)}
\end{equation}

\begin{equation}\label{59}
J_1^ \pm  (x - d,t) = \int\limits_0^\infty  {d\omega }
\frac{{e^{i(\omega  - \Omega  + i\Gamma _ \pm  )t}  - 1}} {{\omega
- \Omega  + i\Gamma _ \pm  }}e^{ - i\frac{\omega } {{v_g }}\left(
{x - d + v_g t} \right)}
\end{equation}

\begin{equation}\label{60}
J_2 (x,t) = \int\limits_0^\infty  {d\omega } \frac{{e^{i(\omega  -
\omega _S )t}  - 1}} {{\omega  - \omega _S }}e^{ - i\frac{\omega }
{{v_g }}\left( {x + v_g t} \right)}
\end{equation}

\begin{equation}\label{61}
J_2 (x - d,t) = \int\limits_0^\infty  {d\omega }
\frac{{e^{i(\omega  - \omega _S )t}  - 1}} {{\omega  - \omega _S
}}e^{ - i\frac{\omega } {{v_g }}\left( {x - d + v_g t} \right)}
\end{equation}

The integrals $J_1^{\pm}(x,t)$ and $J_2(x,t)$ have been calculated
in \cite{Green2023}.
\begin{equation}\label{62}
\begin{gathered}
  J_1^ \pm  (x,t) = e^{ - i\Omega _ \pm  \;t} e^{i\frac{{\left| x \right|}}
{{v_g }}\Omega _ \pm  } E_1\left( {\frac{{\left| x \right|}} {{v_g
}}\Omega _ \pm  } \right) + 2\pi ie^{ - i\frac{{\Omega _ \pm  }}
{{v_g }}(x + v_g t)}  \hfill \\
   - e^{ - i\frac{{x + v_g t}}
{{v_g }}\Omega _ \pm  } E_1\left( { - i\frac{{x + v_g t}}
{{v_g }}\Omega _ \pm  } \right)\; \hfill \\
\end{gathered}
\end{equation}
\begin{equation}\label{63}
\begin{gathered}
  J_2 (x,t) = e^{ - i\frac{{\omega _s }}
{{v_g }}\left( {x + v_g t} \right)} \left( {2\pi i -
\,\rm{ci}\left( {\omega _s \frac{{\left| x \right|}} {{v_g }}}
\right) + \emph{i}\,\rm{si}\left( {\omega _s \frac{{\left| x
\right|}}
{{v_g }}} \right)} \right. \hfill \\
  \left. { + \rm{ci}\left( {\omega _s \frac{{x + v_g t}}
{{v_g }}} \right) + i\,\rm{si}\left( {\omega _s \frac{{x + v_g t}}
{{v_g }}} \right)} \right) \hfill \\
\end{gathered}
\end{equation}
where $\Omega_{\pm}=\Omega-i\Gamma_{\pm}$, $x<0$, and $x+v_g t>0$.

 The integrals $J_1^{\pm}(x-d,t)$ and $J_2(x-d,t)$ can be
obtained from (\ref{62}) and (\ref{63}) simply by replacing $x$
with $x-d$ in corresponding integrals.

\begin{equation}\label{62d}
\begin{gathered}
  J_1^ \pm  (x-d,t) = e^{ - i\Omega _ \pm  \;t} e^{i\frac{{\left| x-d \right|}}
{{v_g }}\Omega _ \pm  } E_1\left( {\frac{{\left| x-d \right|}}
{{v_g }}\Omega _ \pm  } \right)\hfill \\ + 2\pi ie^{ -
i\frac{{\Omega _ \pm }} {{v_g }}(x-d + v_g t)}
   - e^{ - i\frac{{x-d + v_g t}}
{{v_g }}\Omega _ \pm  } E_1\left( { - i\frac{{x-d + v_g t}}
{{v_g }}\Omega _ \pm  } \right)\; \hfill \\
\end{gathered}
\end{equation}
\begin{equation}\label{63d}
\begin{gathered}
  J_2 (x-d,t)\hfill\\ = e^{ - i\frac{{\omega _s }}
{{v_g }}\left( {x-d + v_g t} \right)} \left( {2\pi i -
\,\rm{ci}\left( {\omega _s \frac{{\left| x-d \right|}} {{v_g }}}
\right) + \emph{i}\,\rm{si}\left( {\omega _s \frac{{\left| x-d
\right|}}
{{v_g }}} \right)} \right. \hfill \\
  \left. { + \rm{ci}\left( {\omega _s \frac{{x-d + v_g t}}
{{v_g }}} \right) + \emph{i}\,\rm{si}\left( {\omega _s \frac{{x-d
+ v_g t}}
{{v_g }}} \right)} \right) \hfill \\
\end{gathered}
\end{equation}

First, we investigate the equation (\ref{57}) for $k_{\Omega}d\neq
n\pi$ when $t\rightarrow\infty$. In this case, the quantities
$J_1(x,t), J_1(x-d,t)$ tend to zero exponentially and in the
quantities $J_2(x,t), J_2(x-d,t)$ the time-dependent terms also
tend to zero as the inverse powers of $t$. Therefore, for
$t\rightarrow\infty$ we obtain from (\ref{57}):

\begin{widetext}
\begin{equation}\label{65}
\begin{gathered}
  v(x,t \to \infty ) = e^{ - i\frac{{\omega _s }}
{{v_g }}\left( {x + v_g t} \right)} \frac{1} {2}\sqrt
{\frac{\Gamma } {{4\pi }}} \left( {C_ +   + C_ -  } \right)\left(
{2\pi i - \,\rm{ci}\left( {\omega _s \frac{{\left| x \right|}}
{{v_g }}} \right) + \emph{i}\,\rm{si}\left( {\omega _s
\frac{{\left| x \right|}}
{{v_g }}} \right)} \right) \hfill \\
   + e^{ - i\frac{{\omega _s }}
{{v_g }}\left( {x + v_g t} \right)} \frac{1} {2}\sqrt
{\frac{\Gamma } {{4\pi }}} e^{ik_{\omega _S } d} \left( {C_ +   -
C_ -  } \right)\left( {2\pi i - \,\rm{ci}\left( {\omega _s \left|
{\frac{{x - d}} {{v_g }}} \right|} \right) +
\emph{i}\,\rm{si}\left( {\omega _s \left| {\frac{{x - d}}
{{v_g }}} \right|} \right)} \right) \hfill \\
\end{gathered}
\end{equation}
where $x<0$.
\end{widetext}

The reflectance is obtained from (\ref{65}) in the limit
$x\rightarrow -\infty$:
\begin{equation}\label{66}
\begin{gathered}
  \frac{{\left| {v(x \to -\infty ,t \to \infty )} \right|^2 }}
{{A^2 }} \hfill \\
   = \frac{1}
{{A^2 }}\frac{{\Gamma \pi }} {4}\left| {\left( {C_ +  (1 +
e^{i\frac{{\omega _s }} {{v_g }}d} ) + C_ -  (1 -
e^{i\frac{{\omega _s }}
{{v_g }}d} )} \right)} \right|^2  \hfill \\
\end{gathered}
\end{equation}

Using the explicit expressions for $C_{\pm}$ (\ref{20}) it is not
difficult to show that in resonance, $\omega_S=\Omega$, far from
the first qubit the reflected field is just the reflected incoming
wave,
$v(x\rightarrow-\infty,t\rightarrow\infty)/A=-\exp^{-ik_{\omega_S}(x+v_g
t)}$.

Below we compare (\ref{66}) with non-Markovian reflection
(\ref{RFl}) \cite{Green2015, Liao2015}:

\begin{figure}
  \includegraphics[width=8 cm]{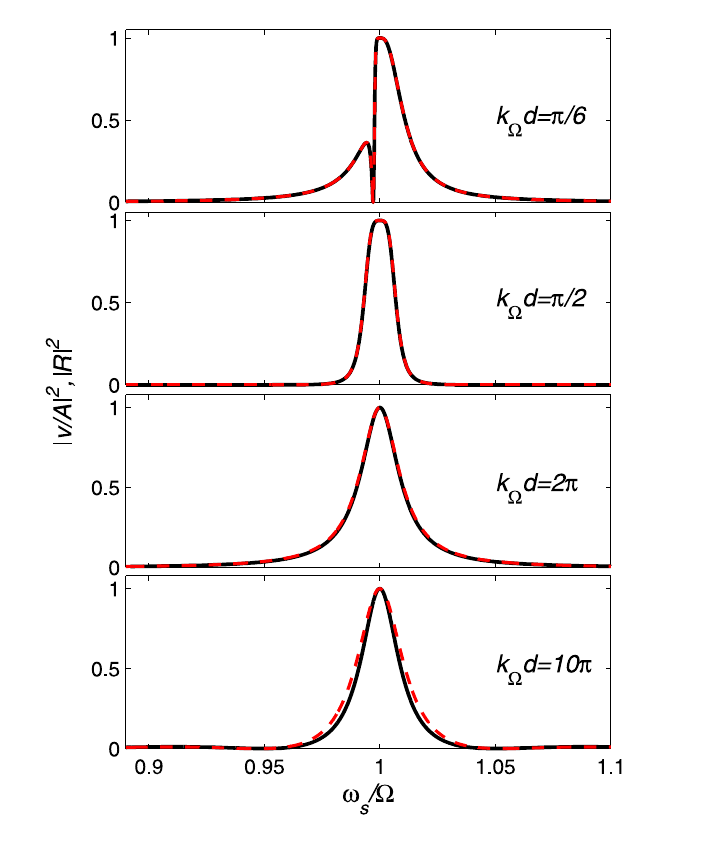}\\
  \caption{The comparison of the Markovian reflectance
  (\ref{66}), (black solid line)
  with non-Markovian expression (\ref{RFl}) (dashed red line).
  $\Gamma/\Omega=0.01$, $\Omega/2\pi=5$GHz.}\label{Fig4}
\end{figure}

\begin{figure}
  \includegraphics[width=8 cm]{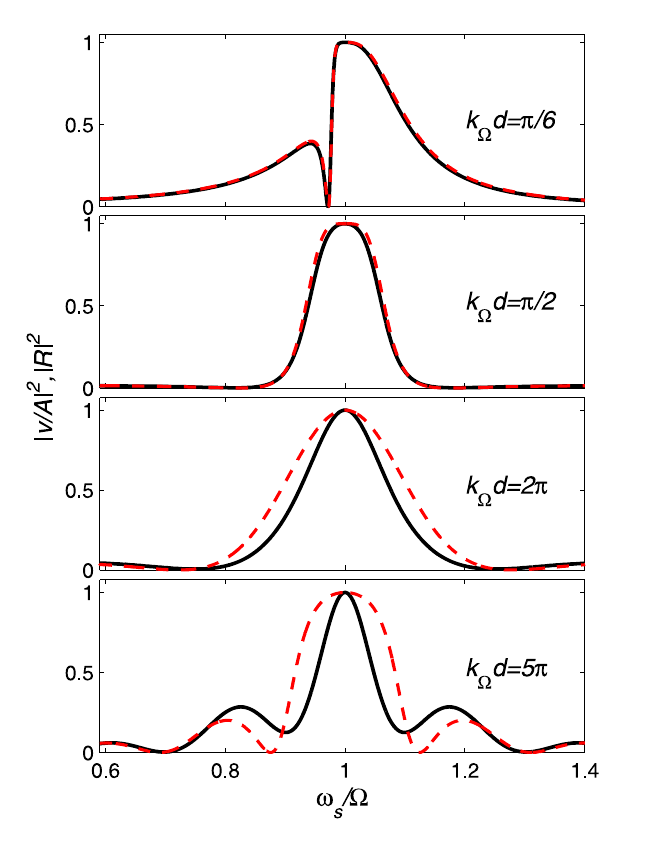}\\
  \caption{The comparison of the Markovian reflectance
  (\ref{66}), (black solid line)
  with non-Markovian expression (\ref{RFl}) (dashed red line).
  $\Gamma/\Omega=0.1$, $\Omega/2\pi=5$GHz.}\label{Fig5}
\end{figure}

\begin{equation}\label{RFl}
|R|^2  = \Gamma ^2 \left| {\frac{{\left( {\omega  - \Omega }
\right)\cos k_\omega  d + \frac{\Gamma } {2}\sin k_\omega  d}}
{{\left( {\omega  - \Omega  + i\frac{\Gamma } {2}} \right)^2  +
\frac{{\Gamma ^2 }} {4}e^{2ik_\omega  d} }}} \right|^2
\end{equation}

The results of the comparison are shown in Fig.\ref{Fig4} and
Fig.\ref{Fig5}. For $\Gamma/\Omega=0.01$ the frequency dependence
(\ref{66}) practically coincides with that of (\ref{RFl}) up to
 $k_{\Omega}d=10\pi$. However, as the value of $\Gamma$
increases, the departure between two transmittances becomes
significant for $k_{\Omega}d=5\pi$ (see Fig.\ref{Fig5}).

For finite $x$ the reflectance at resonance can be obtained
directly from (\ref{65}):

\begin{equation}\label{66a}
\begin{gathered}
  \frac{{\left| {v(x,t \to \infty )} \right|^2 }}
{{A^2 }} = 1 + \frac{1} {\pi }\rm{si}\left( {\Omega \frac{{\left|
\emph{x} \right|}}
{{v_g }}} \right) \hfill \\
   + \frac{1}
{{4\pi ^2 }}\left( {\rm{ci}^2 \left( {\Omega \frac{{\left|
\emph{x} \right|}} {{v_g }}} \right) + si^2 \left( {\Omega
\frac{{\left| \emph{x} \right|}}
{{v_g }}} \right)} \right) \hfill \\
\end{gathered}
\end{equation}

\begin{figure}
  \includegraphics[width=8 cm]{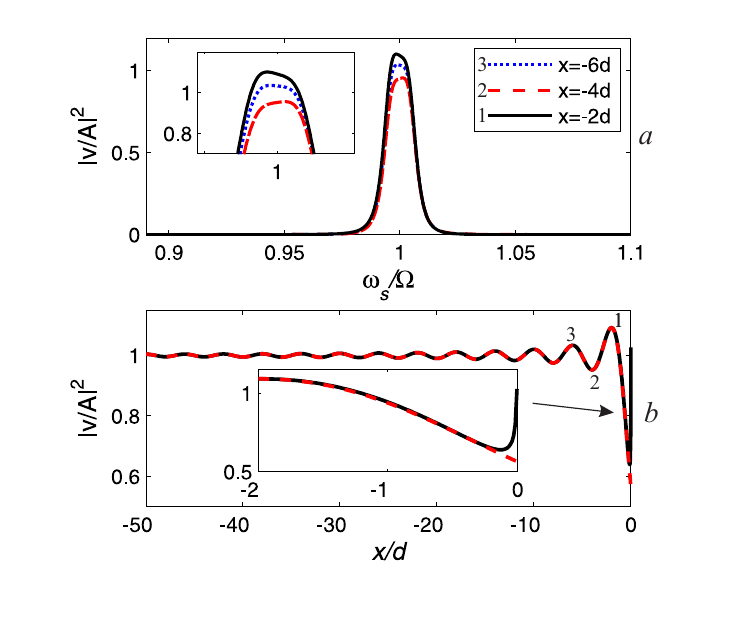}\\
  \caption{Dependence of the peak value of the reflected resonance line on the
  distance $x$ from the first qubit. (a) Resonance lines calculated from (\ref{65})
  for $k_{\Omega}d=\pi/2$ for three successive distances, $x=-2d, x=-4d, x=-6d$. The inset shows the enlarged view of
  resonance lines near the peaks. (b) The dependence of the peak
  value on $x$. The peak values which correspond to those in panel (a) are labelled by
  1, 2, 3, respectively. The inset illustrates the absence of the influence of
  the divergence of $\rm{ci(k_{\Omega}x)}$ at small $x$ on the peak values
  shown in  the panel (a). $k_{\Omega}d=\pi/2$, $\Omega/2\pi=5$GHz, $\Gamma/\Omega=0.01$,
  $d=0.015$m.}\label{Fig8}
\end{figure}

As is seen from (\ref{66a}) the peak value of resonance line
depends on specific $x$-position. This is shown in Fig.\ref{Fig8},
upper panel, where three resonance lines calculated from
(\ref{65}) are shown for $x=-2d, -4d, -6d$, respectively. The
dependence of the peak value on $x$ calculated from (\ref{66a}) is
shown in the lower panel in Fig.\ref{Fig8}. We see that at some
$x$ the peak value of resonance line may exceed unity (black solid
line in Fig.\ref{Fig8}, panel (a)). It is not related to the
divergence of $\rm{ci(k_{\Omega}x)}$ at small $x$ as is shown in
the inset in Fig.\ref{Fig8}, panel (b) where the dashed red line
is calculated from (\ref{66a}) with $\rm{ci(k_{\Omega}x)}$ being
neglected. The peak value which is higher than unity at a single
frequency and a single spatial point does not contradict the
energy conservation, which should account for the whole frequency
range and whole volume. The origin of the higher-than-unity peak
comes mainly from the interference term
$(1/\pi)\rm{si}(\Omega|x|/v_g)$ in (\ref{66a}).

Next, we analyze the reflected field for $k_{\Omega}d=n\pi$, where
$n$ is the integer. For even $n$ the quantities $C_{\pm}$ are
given in (\ref{48}) and we obtain for reflected wave the
expression which is similar to (\ref{49}):

\begin{equation}\label{67}
\begin{gathered}
  v(x,t) = \hfill\\ - \frac{1}
{2}\sqrt {\frac{\Gamma }
{{4\pi }}} C_ +  \left( {J_1  (x,t) + J_1  (x - d,t) - J_2 (x,t) - J_2 (x - d,t)} \right) \hfill \\
   - \frac{1}
{2}\sqrt {\frac{\Gamma }
{{4\pi }}} C_ -  \left( {J_3 (x,t) - J_3 (x - d,t) - J_2 (x,t) + J_2 (x - d,t)} \right) \hfill \\
\end{gathered}
\end{equation}
where
\begin{equation}\label{68}
J_1^{} (x,t) = \int\limits_0^\infty  {d\omega } \frac{{e^{i(\omega
- \Omega  + i\Gamma )t}  - 1}} {{\omega  - \Omega  + i\Gamma }}e^{
- i\frac{\omega } {{v_g }}\left( {x + v_g t} \right)}
\end{equation}

\begin{equation}\label{69}
J_2 (x,t) = \int\limits_0^\infty  {d\omega } \frac{{e^{i(\omega  -
\omega _S )t}  - 1}} {{\omega  - \omega _S }}e^{ - i\frac{\omega }
{{v_g }}\left( {x + v_g t} \right)}
\end{equation}

\begin{equation}\label{70}
J_3^{} (x,t) = \int\limits_0^\infty  {d\omega } \frac{{e^{i(\omega
- \Omega )t}  - 1}} {{\omega  - \Omega }}e^{ - i\frac{\omega }
{{v_g }}\left( {x + v_g t} \right)}
\end{equation}

The expressions for $J_1(x-d,t), J_2(x-d,t), J_3(x-d,t)$ can be
obtained from equations (\ref{68}), (\ref{69}), (\ref{70}) by
replacing $x$ with $x-d$ in corresponding equations.

The quantities $J_1(x,t), J_1(x-d,t)$ in (\ref{67}) are given by
the expressions (\ref{62}) and (\ref{62d}), where $\Omega_{\pm}$
is replaced by $\Omega-i\Gamma$. The quantities $J_2(x,t),
J_2(x-d,t)$ are given in (\ref{63}) (\ref{63d}),  and the
quantities $J_3(x,t), J_3(x-d,t)$ are given by the expressions
(\ref{63}), (\ref{63d}) where $\omega_S$ is replaced by $\Omega$.

We investigate the equation (\ref{67}) when $t\rightarrow\infty$.
In this case, the quantities $J_1(x,t), J_1(x-d,t)$ tend to zero
exponentially and in the quantities $J_2(x,t), J_2(x-d,t)$,
$J_3(x,t), J_3(x-d,t)$ the time-dependent terms also tend to zero
as the inverse powers of $t$. Therefore, for $t\rightarrow\infty$
we obtain from (\ref{67}):

\begin{widetext}
\begin{equation}\label{71}
\begin{gathered}
  v(x,t \to \infty ) = e^{ - i\frac{{\omega _s }}
{{v_g }}\left( {x + v_g t} \right)} \frac{1} {2}\sqrt
{\frac{\Gamma } {{4\pi }}} \left( {C_ +   + C_ -  } \right)\left(
{2\pi i - \,\rm{ci}\left( {\omega _s \frac{{\left| x \right|}}
{{v_g }}} \right) + \emph{i}\,\rm{si}\left( {\omega _s
\frac{{\left| x \right|}}
{{v_g }}} \right)} \right) \hfill \\
   + e^{ - i\frac{{\omega _s }}
{{v_g }}\left( {x + v_g t} \right)} \frac{1} {2}\sqrt
{\frac{\Gamma } {{4\pi }}} e^{ik_{\omega _S } d} \left( {C_ +   -
C_ -  } \right)\left( {2\pi i - \,\rm{ci}\left( {\omega _s \left|
{\frac{{x - d}} {{v_g }}} \right|} \right) +
\emph{i}\,\rm{si}\left( {\omega _s \left| {\frac{{x - d}}
{{v_g }}} \right|} \right)} \right) \hfill \\
   + e^{ - i\frac{\Omega }
{{v_g }}\left( {x + v_g t} \right)} \frac{1} {2}\sqrt
{\frac{\Gamma } {{4\pi }}} C_ -  \left( {\,\rm{ci}\left( {\Omega
\frac{{\left| x \right|}} {{v_g }}} \right) -
\emph{i}\,\rm{si}\left( {\Omega \frac{{\left| x \right|}} {{v_g
}}} \right) - \,\rm{ci}\left( {\Omega \frac{{\left| {x - d}
\right|}} {{v_g }}} \right) + \emph{i}\,\rm{si}\left( {\Omega
\frac{{\left| {x - d} \right|}}
{{v_g }}} \right)} \right) \hfill \\
\end{gathered}
\end{equation}
where $x<0$, $k_{\Omega}d=2n\pi$. From (\ref{71}) we obtain for
$\omega_S=\Omega$ the equation which describes the dependence of
the peak value of the reflected resonace line on the position
point $x$.

\begin{equation}\label{71a}
\begin{gathered}
  \frac{{\left| {v(x,t \to \infty )} \right|^2 }}
{{A^2 }} = 1 + \frac{1} {{2\pi }}\rm{si}\left( {2\pi \frac{{\left|
\emph{x} \right|}} {\emph{d}}} \right) + \frac{1} {{2\pi
}}si\left( {2\pi \frac{{\left| {\emph{x} - \emph{d}} \right|}}
{\emph{d}}} \right) \hfill \\
   + \frac{1}
{{4\pi ^2 }}\left[ {\left( {\rm{ci}\left( {2\pi \frac{{\left|
\emph{x} \right|}} {\emph{d}}} \right) + ci\left( {2\pi
\frac{{\left| {\emph{x} - \emph{d}} \right|}} {\emph{d}}} \right)}
\right)^2  + \left( {\rm{si}\left( {2\pi \frac{{\left| \emph{x}
\right|}} {\emph{d}}} \right) + si\left( {2\pi \frac{{\left|
{\emph{x} - \emph{d}} \right|}}
{\emph{d}}} \right)} \right)^2 } \right] \hfill \\
\end{gathered}
\end{equation}
where where $x<0$, $k_{\Omega}d=2\pi$.
\end{widetext}

If  $x\rightarrow\infty$, all sine and cosine integrals in
(\ref{71}) tend to zero and we obtain for the reflectance the
equation (\ref{66}) where $C_{\pm}$ are given in (\ref{48}).

\begin{figure}
  \includegraphics[width=8 cm]{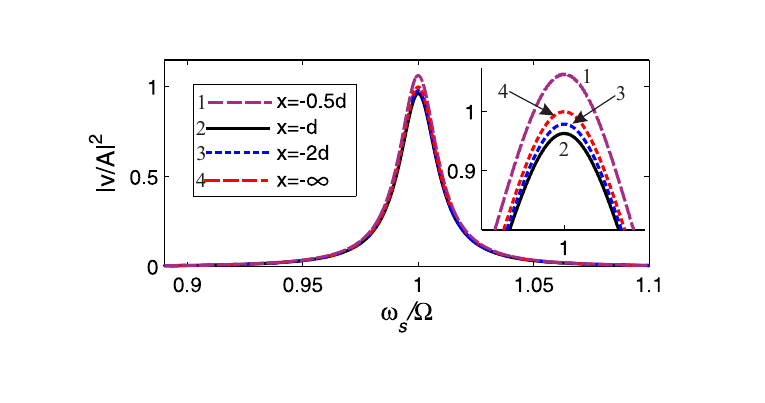}\\
  \caption{Reflected resonance lines for $k_{\Omega} d=2\pi$ and several $x$, calculated
  from (\ref{71}). (1) $x=-0.5$, lilac dashed line;  (2) $x=-d$, black solid line; (3) $x=-2d$,
  blue dashed line;
  (4) $x=-\infty$, red dashed line. For clarity, the upper part of resonance lines
  is shown in the inset.
  $\Gamma/\Omega=0.01$, $\Omega/2\pi=5$GHz, $d=0.06$m.}\label{Fig10}
\end{figure}
Reflected resonance lines for $k_{\Omega} d=2\pi$ and several $x$,
calculated from (\ref{71}) for $t=5\times 10^{-6}$c are shown in
Fig.\ref{Fig10}. A small amplification of the resonance peak (line
1 in Fig.\ref{Fig10}) is due to the influence of the interference
term in the first line in (\ref{71a}). Similar effect is known in
Fabry-Perot interferometer with semitransparent mirrors where the
strength mode of electric field can significantly exceed the
amplitude of the input field \cite{Ley1987}. This amplification
effect will be discussed in more detail at the end of this
section.

It is not difficult to perform similar calculations for
$k_{\Omega}d=n\pi$ if $n$ is odd number. For this case we obtain
for $v(x,t)$ the expression which is similar to (\ref{67}).

\begin{equation}\label{72}
    \begin{gathered}
  v (x,t) = \hfill\\ - \frac{1}
{2}\sqrt {\frac{\Gamma }
{{4\pi }}} C_ +  \left( {J_3  (x,t) + J_3  (x - d,t) - J_2 (x,t) - J_2 (x - d,t)} \right) \hfill \\
   - \frac{1}
{2}\sqrt {\frac{\Gamma }
{{4\pi }}} C_ -  \left( {J_1  (x,t) - J_1  (x - d,t) - J_2 (x,t) + J_2 (x - d,t)} \right) \hfill \\
\end{gathered}
\end{equation}
where $C_{\pm}$ are given in (\ref{56}).

In the limit   $t\rightarrow\infty$ we obtain from (\ref{72}) the
expression which is analogous to equation (\ref{71}) with the only
exception: in the third line in (\ref{71}) $C_-$ should be
replaced with $C_+$.

As in the $n$-even case, the reflectance here is also given by the
expression (\ref{66}).

\subsection{Scattering waves between the qubits}

Between qubits, $0<x<d$, the photon field is a superposition of
the forward and backward travelling waves.
\begin{equation}\label{73}
    w(x,t)=u_0(x,t)+u_1(x,t)+v(x,t)
\end{equation}

The general expressions for $u_1(x,t)$ and $v(x,t)$ are given in
(\ref{35}) and (\ref{57}) where $0<x<d$. First we consider the
case $k_{\Omega}d\neq n\pi$, so that in expressions (\ref{35}) and
(\ref{57}) we neglect the terms $I_1^{\pm}(x,t)$,
$I_1^{\pm}(x-d,t)$, and $J_1^{\pm}(x,t)$, $J_1^{\pm}(x-d,t)$ which
tend exponentially to zero as $t\rightarrow\infty$.

\begin{equation}\label{74}
\begin{gathered}
  u_1 (x,t) = \frac{1}
{2}\sqrt {\frac{\Gamma }
{{4\pi }}} \left( {C_ +   + C_ -  } \right)I_2 (x,t) \hfill \\
   + \frac{1}
{2}\sqrt {\frac{\Gamma }
{{4\pi }}} \left( {C_ +   - C_ -  } \right)I_2 (x - d,t) \hfill \\
\end{gathered}
\end{equation}

\begin{equation}\label{75}
\begin{gathered}
  v(x,t) =  \frac{1}
{2}\sqrt {\frac{\Gamma }
{{4\pi }}} \left( {C_ +   + C_ -  } \right)J_2 (x,t) \hfill \\
   + \frac{1}
{2}\sqrt {\frac{\Gamma }
{{4\pi }}} \left( {C_ +   - C_ -  } \right)J_2 (x - d,t) \hfill \\
\end{gathered}
\end{equation}
In these equations $x>0$, and $x-d<0$.

In (\ref{74}) the quantity $I_2(x,t)$ where $x>0$ is given in
(\ref{42}). The calculation of $I_2(x-d,t)$ where $x-d<0$ provides
the following result (see expression (\ref{B16}) in the Appendix
B):
\begin{equation}\label{76}
\begin{gathered}
  I_2 (x - d,t)\hfill\\ = e^{i\frac{{\omega _s }}
{{v_g }}\left( {x - d - v_g t} \right)} \left\{ { -
\,\rm{ci}\left( {\omega _s \left| {\frac{{x - d}} {{v_g }}}
\right|} \right) - \emph{i}\,\rm{si}\left( {\omega _s \left|
{\frac{{x - d}}
{{v_g }}} \right|} \right)} \right. \hfill \\
  \left. { + \,\rm{ci}\left( {\omega _s \left| {\frac{{x - d - v_g t}}
{{v_g }}} \right|} \right) + \emph{i}\,\rm{si}\left( {\omega _s
\left| {\frac{{x - d - v_g t}}
{{v_g }}} \right|} \right)} \right\} \hfill \\
\end{gathered}
\end{equation}
Next, in the expressions (\ref{42}) and (\ref{76}) we neglect the
terms which are tend to zero as the inverse powers of $t$.
Therefore, for interqubit forward travelling wave we obtain:
\begin{widetext}
\begin{equation}\label{77}
\begin{gathered}
  u (x,t\rightarrow\infty) = e^{i\frac{{\omega _s }}
{{v_g }}\left( {x - v_g t} \right)} \left[ {A + \frac{1} {2}\sqrt
{\frac{\Gamma } {{4\pi }}} \left( {C_ +   + C_ -  } \right)\left(
{2\pi i - \,\rm{ci}\left( {\omega _s \frac{x} {{v_g }}} \right) +
\emph{i}\,\rm{si}\left( {\omega _s \frac{x}
{{v_g }}} \right)} \right)} \right] \hfill \\
   - e^{i\frac{{\omega _s }}
{{v_g }}\left( {x - d - v_g t} \right)} \frac{1} {2}\sqrt
{\frac{\Gamma } {{4\pi }}} \left( {C_ +   - C_ -  } \right)\left(
{  \,\rm{ci}\left( {\omega _s \left| {\frac{{x - d}} {{v_g }}}
\right|} \right) + \emph{i}\,\rm{si}\left( {\omega _s \left|
{\frac{{x - d}}
{{v_g }}} \right|} \right)} \right) \hfill \\
\end{gathered}
\end{equation}
where $0<x<d$.
\end{widetext}
Now we calculate the interqubit backward travelling wave. In
equation (\ref{75}) the quantity $J_2(x-d,t))$ is given by the
equation (\ref{63d}). The calculation of $J_2(x,t)$ for $x>0$
yields the following result (see expression (\ref{B20}) in the
Appendix B):

\begin{equation}\label{78}
\begin{gathered}
  J_2 (x,t) =  - e^{ - i\frac{{\omega _s }}
{{v_g }}\left( {x + v_g t} \right)} \left( {\,\rm{ci}\left(
{\omega _s \frac{{ x }} {{v_g }}} \right) +
\emph{i}\,\rm{si}\left( {\omega _s \frac{{ x }}
{{v_g }}} \right)} \right. \hfill \\
  \left. { - \rm{ci}\left( {\omega _s \frac{{x + v_g t}}
{{v_g }}} \right) - \emph{i}\,\rm{si}\left( {\omega _s \frac{{x +
v_g t}}
{{v_g }}} \right)} \right) \hfill \\
\end{gathered}
\end{equation}

In the expressions (\ref{63d}) and (\ref{78}) we neglect the terms
which are tend to zero as the inverse powers of $t$. Therefore,
for interqubit backward travelling wave we obtain:
\begin{widetext}
\begin{equation}\label{79}
\begin{gathered}
  v(x,t\rightarrow\infty) = -e^{ - i\frac{{\omega _s }}
{{v_g }}\left( {x + v_g t} \right)} \frac{1} {2}\sqrt
{\frac{\Gamma } {{4\pi }}} \left( {C_ +   + C_ -  } \right)\left(
{\rm{ci}\left( {\omega _s \frac{{ x }} {{v_g }}} \right) +
\emph{i}\,\rm{si}\left( {\omega _s \frac{{ x }}
{{v_g }}} \right)} \right) \hfill \\
   + e^{ - i\frac{{\omega _s }}
{{v_g }}\left( {x - d + v_g t} \right)} \frac{1} {2}\sqrt
{\frac{\Gamma } {{4\pi }}} \left( {C_ +   - C_ -  } \right)\left(
{2\pi i - \,\rm{ci}\left( {\omega _s \frac{{\left| {x - d}
\right|}} {{v_g }}} \right) + \emph{i}\,\rm{si}\left( {\omega _s
\frac{{\left| {x - d} \right|}}
{{v_g }}} \right)} \right) \hfill \\
\end{gathered}
\end{equation}
where $0<x<d$.
\end{widetext}
From (\ref{77}) and (\ref{79}) we obtain for the interqubt field
energy $|u+v|^2$  at the resonance point $\omega_S=\Omega$ the
following simple expression:
\begin{equation}\label{interf}
\begin{gathered}
  \frac{{\left| {u(x,t \to \infty ) + v(x,t \to \infty )} \right|^2 }}
{{A^2 }} \hfill \\
   = \frac{1}{\pi ^2 }\left( {\rm{ci}(k_\Omega  x)\cos \left( {k_\Omega  x} \right) + \rm{si}(k_\Omega  x)\sin \left( {k_\Omega  x} \right)} \right)^2  \hfill \\
\end{gathered}
\end{equation}
where $k_{\Omega}=\Omega/v_g$, $0<x<d$.

\begin{figure}
  \includegraphics[width=8 cm]{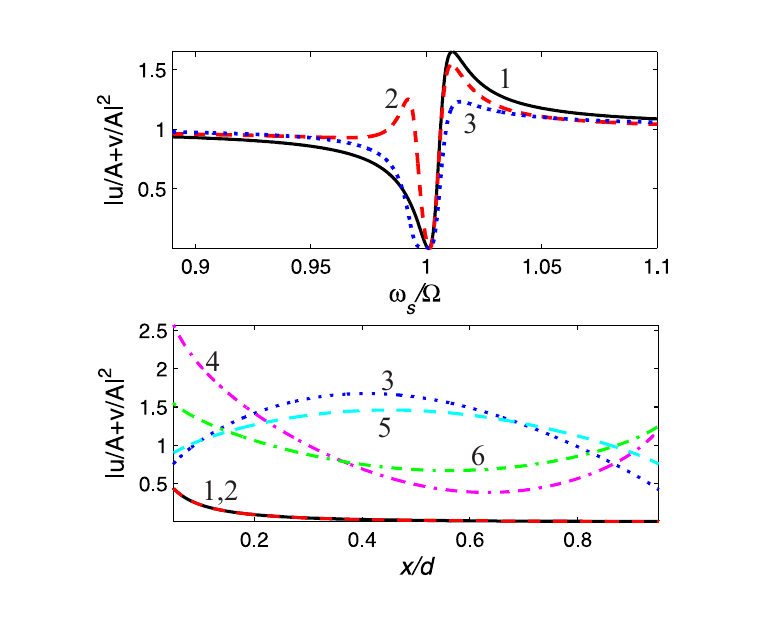}\\
  \caption{Photon field between qubits for $k_{\Omega}d=\pi/2$. Upper panel. Interqubit resonance line
  $|u+v|^2/A^2$ calculated from (\ref{77}) and (\ref{79}) for three spatial points.
  $x=0.5d$ (1;black solid line), $x=0.25d$ (2; red dashed line), $x=0.75d$
   (3;dashed blue line). Lower panel. Dependence of the field energy $|u+v|^2/A^2$
   on $x$ for several values of the photon frequency. (1) $\omega_S=\Omega$
    and (2) $\omega_S=\Omega$ calculated from (\ref{interf})), (3) $\omega_S=1.01\Omega$;
     (4) $\omega_S=0.99\Omega$; (5) $\omega_S=1.02$; (6) $\omega_S=0.98\Omega$.
     $\Gamma/\Omega=0.01$, $\Omega/2\pi=5$GHz, $d=0.015$m.}\label{Fig9}
\end{figure}

The behavior of the photon field between qubits calculated from
(\ref{77}) and (\ref{79}) is shown in Fig.\ref{Fig9} for
$k_{\Omega}d=\pi/2$. In the upper panel the resonance lines
  $|u+v|^2/A^2$ are shown for three spatial points,
  $x=0.5d$, $x=0.25d$, and $x=0.75d$.
   In the lower panel the dependence of the field energy $|u+v|^2/A^2$
   on $x$ is shown for several values of the photon frequency,
   $\omega_S=\Omega$,
    $\omega_S=1.01\Omega$,
     $\omega_S=0.99\Omega$, $\omega_S=1.02$; $\omega_S=0.98\Omega$.
     $\Gamma/\Omega=0.01$, $\Omega/2\pi=5$GHz, $d=0.015$m.
In Fig.\ref{Fig9} we observe a significant amplification of the
field energy for some spatial points and some frequencies. We
discuss this effect in a more detail in the end of this section.

\subsubsection{Interqubit field for $k_{\Omega}d=n\pi$}

Below for definiteness we take $k_{\Omega}d=n\pi$ where $n$ is
even number. In this case, the interqubit forward and backward
travelling waves, $u_1(x,t)$ and $v(x,t)$ are given by the
equations (\ref{49}), (\ref{67}) where $x>0$ and $x-d<0$, and
$C_{\pm}$ are defined in (\ref{48}). Neglecting in these equations
the exponential decaying terms we obtain:
\begin{equation}\label{80}
\begin{gathered}
  u_1 (x,t) = \frac{1}
{2}\sqrt {\frac{\Gamma }
{{4\pi }}} \left( {C_ +   + C_ -  } \right)I_2 (x,t) \hfill \\
   + \frac{1}
{2}\sqrt {\frac{\Gamma }
{{4\pi }}} \left( {C_ +   - C_ -  } \right)I_2 (x - d,t) \hfill \\
-\frac{1}{2}\sqrt {\frac{\Gamma } {{4\pi
}}}C_-\left(I_3(x,t)-I_3(x-d,t)\right)
\end{gathered}
\end{equation}

\begin{equation}\label{81}
\begin{gathered}
  v(x,t) =  \frac{1}
{2}\sqrt {\frac{\Gamma }
{{4\pi }}} \left( {C_ +   + C_ -  } \right)J_2 (x,t) \hfill \\
   + \frac{1}
{2}\sqrt {\frac{\Gamma }
{{4\pi }}} \left( {C_ +   - C_ -  } \right)J_2 (x - d,t) \hfill \\
-\frac{1}{2}\sqrt {\frac{\Gamma } {{4\pi
}}}C_-\left(J_3(x,t)-J_3(x-d,t)\right)
\end{gathered}
\end{equation}
In (\ref{80}) the quantities $I_2(x,t)$ where $x>0$, and
$I_2(x-d,t)$ where $x-d<0$ are given in (\ref{42}) and (\ref{76}),
respectively. The integrals $I_3(x,t)$ where $x>0$ and
$I_3(x-d,t)$ where $x-d<0$ are given by the expressions
(\ref{B15}) and (\ref{B16}), respectively, where $\omega_S$ should
be replaced with $\Omega$.

\begin{equation}\label{82}
 \begin{gathered}
    I_3 (x,t) = e^{i\Omega
    T}\hfill\\\times\left(-\rm{ci}(\Omega\tau)+\emph{i}\,\rm{si}(\Omega\tau)+\rm{ci}(\Omega|T|)+\emph{i}\,\rm{si}(\Omega|T|)\right)\hfill\\
    +2i\pi e^{i\Omega T}
\end{gathered}
\end{equation}
where $\tau=x/v_g>0$, $T=(x-v_gt)/v_g<0$.

\begin{equation}\label{83}
 \begin{gathered}
    I_3 (x-d,t) = e^{i\Omega
    T}\hfill\\\times\left(-\rm{ci}(\Omega|\tau|)-\emph{i}\,\rm{si}(\Omega|\tau|)+\rm{ci}(\Omega|T|)+\emph{i}\,\rm{si}(\Omega|T|)\right)
\end{gathered}
\end{equation}
where $\tau=(x-d)/v_g<0$, $T=(x-d-v_gt)/v_g<0$.

Therefore, when $t\rightarrow\infty$ for interqubit forward
scattering field we obtain:
\begin{widetext}
\begin{equation}\label{84}
\begin{gathered}
  u(x,t \to \infty ) = e^{i\frac{{\omega _s }}
{{v_g }}\left( {x - v_g t} \right)} \left[ {A + \frac{1} {2}\sqrt
{\frac{\Gamma } {{4\pi }}} \left( {C_ +   + C_ -  } \right)\left(
{2\pi i - \,\rm{ci}\left( {\omega _s \frac{x} {{v_g }}} \right) +
\emph{i}\,\rm{si}\left( {\omega _s \frac{x}
{{v_g }}} \right)} \right)} \right] \hfill \\
   + e^{i\frac{{\omega _s }}
{{v_g }}\left( {x - d - v_g t} \right)} \frac{1} {2}\sqrt
{\frac{\Gamma } {{4\pi }}} \left( {C_ +   - C_ -  } \right)\left(
{ - \,\rm{ci}\left( {\omega _s \left| {\frac{{x - d}} {{v_g }}}
\right|} \right) - \emph{i}\,\rm{si}\left( {\omega _s \left|
{\frac{{x - d}}
{{v_g }}} \right|} \right)} \right) \hfill \\
   - e^{i\frac{\Omega }
{{v_g }}\left( {x - v_g t} \right)} \frac{1} {2}\sqrt
{\frac{\Gamma } {{4\pi }}} C_ -  \left( {2\pi i - \rm{ci}\left(
{\Omega \frac{x} {{v_g }}} \right) + \emph{i}\,\rm{si}\left(
{\Omega \frac{x} {{v_g }}} \right) + \rm{ci}\left( {\Omega
\frac{{\left| {x - d} \right|}} {{v_g }}} \right) +
\emph{i}\,\rm{si}\left( {\Omega \frac{{\left| {x - d} \right|}}
{{v_g }}} \right)} \right) \hfill \\
\end{gathered}
\end{equation}
where $0<x<d$, $k_{\Omega}d=n\pi$.
\end{widetext}

In equation (\ref{81}) which describes the backward scattering
field between qubits the integral $J_2(x,t)$ where $x>0$ is given
in (\ref{78}), the integral $J_2(x-d,t)$ is given by the equation
(\ref{63d}), and the integral $J_3(x,t)$ is given by the
expression (\ref{B20}) where $\omega_S$ should be replaced with
$\Omega$:

\begin{equation}\label{86}
\begin{gathered}
 J_3 (x,t) = e^{-i\Omega
    T}\hfill\\\times\left(-\rm{ci}(\Omega\tau)-\emph{i}\,\rm{si}(\Omega\tau)+\rm{ci}(\Omega T)+\emph{i}\,\rm{si}(\Omega T)\right)
\end{gathered}
\end{equation}
where $\tau=x/v_g>0$, $T=(x+v_gt)/v_g>0$.

The integral $J_3(x-d,t)$ is obtained from the expression
(\ref{B19}) where $x$ should be replaced with $x-d$, and
$\omega_S$ should be replaces with $\Omega$:
\begin{equation}\label{87}
 \begin{gathered}
    J_3 (x-d,t) = e^{-i\Omega
    T}\hfill\\\times\left(-\rm{ci}(\Omega|\tau|)+\emph{i}\,\rm{si}(\Omega|\tau|)+\rm{ci}(\Omega T)+\emph{i}\,\rm{si}(\Omega T)\right)\hfill\\
    +2i\pi e^{-i\Omega T}
\end{gathered}
\end{equation}
where $\tau=(x-d)/v_g<0$, $T=(x-d+v_gt)/v_g>0$.

Therefore, when $t\rightarrow\infty$ for interqubit backward
scattering field we obtain:
\begin{widetext}
\begin{equation}\label{88}
\begin{gathered}
  v(x,t \to \infty ) = -e^{ - i\frac{{\omega _s }}
{{v_g }}\left( {x + v_g t} \right)} \frac{1} {2}\sqrt
{\frac{\Gamma } {{4\pi }}} \left( {C_ +   + C_ -  } \right)\left(
{\rm{ci}\left( {\omega _s \frac{x} {{v_g }}} \right) +
\emph{i}\,\rm{si}\left( {\omega _s \frac{x}
{{v_g }}} \right)} \right) \hfill \\
   + e^{ - i\frac{{\omega _s }}
{{v_g }}\left( {x - d + v_g t} \right)} \frac{1} {2}\sqrt
{\frac{\Gamma } {{4\pi }}} \left( {C_ +   - C_ -  } \right)\left(
{ - \,\rm{ci}\left( {\omega _s \frac{{\left| {x - d} \right|}}
{{v_g }}} \right) + \emph{i}\,\rm{si}\left( {\omega _s
\frac{{\left| {x - d} \right|}}
{{v_g }}} \right)} \right) \hfill \\
   + e^{ - i\frac{\Omega }
{{v_g }}\left( {x + v_g t} \right)} \frac{1} {2}\sqrt
{\frac{\Gamma } {{4\pi }}} C_ -  \left( {2\pi i + \rm{ci}\left(
{\Omega \frac{x} {{v_g }}} \right) + \emph{i}\,\rm{si}\left(
{\Omega \frac{x} {{v_g }}} \right) - \,\rm{ci}\left( {\Omega
\frac{{\left| {x - d} \right|}} {{v_g }}} \right) +
\emph{i}\,\rm{si}\left( {\Omega \frac{{\left| {x - d} \right|}}
{{v_g }}} \right)} \right) \hfill \\
\end{gathered}
\end{equation}
where $0<x<d$, $k_{\Omega}d=n\pi$.
\end{widetext}
In equations (\ref{84}) and (\ref{88}) $C_{\pm}$ are given in
(\ref{48}).

Between the qubits the photon field is a superposition of the
forward, (\ref{84}), and backward, (\ref{88}) travelling waves.

\begin{figure}
  \includegraphics[width=8cm]{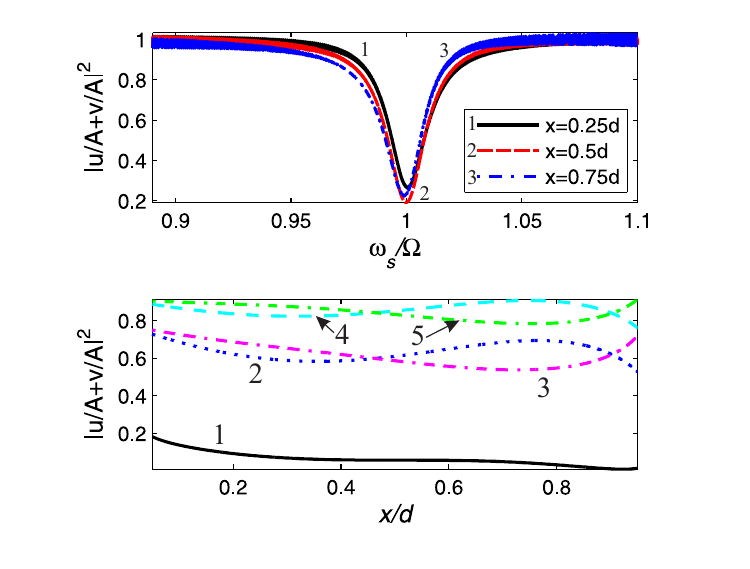}\\
  \caption{Photon field between qubits for $k_{\Omega}d=2\pi$. Upper panel. Interqubit resonance line
  $|u+v|^2/A^2$ calculated from (\ref{84}) and (\ref{88}) for three spatial points.
  $x=0.25d$ (1;black solid line), $x=0.5d$ (2; red dashed line), $x=0.75d$
   (3;dotted-dashed blue line). Lower panel. Dependence of the field energy $|u+v|^2/A^2$
   on $x$ for several values of the photon frequency. (1)
   $\omega_S=\Omega$; (2) $\omega_S=1.01\Omega$;
     (3) $\omega_S=0.99\Omega$; (4) $\omega_S=1.02$; (5) $\omega_S=0.98\Omega$.
     $\Gamma/\Omega=0.01$, $\Omega/2\pi=5$GHz, $d=0.06$m.}\label{Fig11}
\end{figure}

The behavior of the photon field between qubits calculated from
(\ref{84}) and (\ref{88}) for $k_{\Omega}d=2\pi$ and $t=5\times
10^{-6}$c is shown in Fig.\ref{Fig11}. In the upper panel the
resonance lines
  $|u+v|^2/A^2$ are shown for three spatial points,
  $x=0.25d$ (1;black solid line), $x=0.5d$ (2; red dashed line), $x=0.75d$
   (3;dotted-dashed blue line).
   In the lower panel the dependence of the field energy $|u+v|^2/A^2$
   on $x$ is shown for several values of the photon frequency,
   (1)
   $\omega_S=\Omega$; (2) $\omega_S=1.01\Omega$;
     (3) $\omega_S=0.99\Omega$; (4) $\omega_S=1.02$; (5) $\omega_S=0.98\Omega$.

If we compare the forms of resonance lines and the peak values in
Fig.\ref{Fig9} with those in Fig.\ref{Fig11} we observe a drastic
difference. For example, a significant amplification of the field
energy for some spatial points and frequencies is seen in
Fig.\ref{Fig9}. We attribute this difference to the constructive
interference between the radiation of symmetric and asymmetric
states of a two-qubit system. For $k_{\Omega}d=\pi/2$ both these
states are not decoupled from the waveguide photons. If
$k_{\Omega}d=2\pi$ (Fig.\ref{Fig11}) the asymmetric state is
decoupled from the photon field and, therefore, does not
contribute to the radiation.

\section{Conclusion}
In summary, we have developed the time-dependent theory of the
scattering of a narrow single-photon Gaussian pulse from a
two-qubit system embedded in 1D open waveguide. The theory is
valid for Markov approximation when time delay effects between
qubits are neglected. This approximation allows us to obtain
explicit analytical expressions for the forward and backward
travelling waves, their spatial and temporal distribution. We show
that the scattered fields consist of several parts: a free field
of incoming photon, the part which describes a spontaneous
exponential decay of excited qubits, a slowly decaying part dying
out as the inverse powers of $t$, which is the manifestation of
subradiant emission, and a lossless part which represents a steady
state solution as $t\rightarrow\infty$. We systematically compare
the transmittance and reflectance for our model with those for
non-Markovian case which are known from the literature. It turned
out that our transmittance and reflectance work well for
$k_{\Omega}d<10\pi$ if $\Gamma/\Omega<0.01$. If the interqubit
distance $d$ is equal to wavelength or half wavelength of
resonance frequency, the field energy exhibits the beatings with
the detuning frequency $\omega_S-\Omega$.

Our calculations show that spatial effects can persist on the
scale of several $\lambda$'s (see Fig.\ref{Fig6} and
Fig.\ref{Fig8}).  For on-chip realization this length is not small
compared with the dimensions of a superconducting qubit (typically
several microns). The power of microwave signal is so low that the
use of linear amplifiers for the detection of the qubit signal is
a common practice. The current opportunity for on-chip realization
of superconducting qubits with associated circuitry allows for the
placement of the amplifier within the order of the wavelength from
the qubits \cite{Lec2021}. Therefore, in microwave range the
near-field effects can in principle be detectable.

We believe that the results obtained in this paper may have some
practical applications in quantum information technologies
including single-photon detection in a microwave domain as well as
the optimization of the control and readout of a qubit's quantum
state.

\textbf{Acknowledgments}. The authors thank O. V. Kibis  and A. N.
Sultanov for fruitful discussions. The work is supported by the
Ministry of Science and Higher Education of Russian Federation
under the project FSUN-2023-0006.

\textbf{Author contributions}. YSG wrote the manuscript and
contributed to its theoretical interpretation. AAS and AGM
performed analytical calculations and computer simulations. All
authors discussed the results and commented on the manuscript. The
authors declare that they have no competing interests.

\textbf{Data Availability Statement}. The manuscript has no
associated data in a public repository.

\appendix
\section{Derivation of equations for qubits' amplitudes (\ref{12}),
(\ref{13})}

Taking the qubits' amplitudes $\beta_n(t')$ out of integrals in
(\ref{10}), (\ref{11}) we obtain:
\begin{equation}\label{A1}
\begin{gathered}
  \frac{{d\beta _1 }}
{{dt}} =  - i\int\limits_0^\infty  {d\omega } g(\omega )\gamma _0 \left( \omega  \right)e^{ - i(\omega  - \Omega )t}  \hfill \\
   - 2\beta _1 (t)\int\limits_0^\infty  {d\omega } g^2 (\omega )I(\omega ,t) \hfill \\
   - 2\beta _2 (t)\int\limits_0^\infty  {d\omega } g^2 (\omega )\cos (k_\omega  d)I(\omega ,t) \hfill \\
\end{gathered}
\end{equation}

\begin{equation}\label{A2}
\begin{gathered}
  \frac{{d\beta _2 }}
{{dt}} =  - i\int\limits_0^\infty  {d\omega } g(\omega )\gamma _0 \left( \omega  \right)e^{ik_\omega  d} e^{ - i(\omega  - \Omega )t}  \hfill \\
   - 2\beta _1 (t)\int\limits_0^\infty  {d\omega } g^2 (\omega )\cos (k_\omega  d)I(\omega ,t) \hfill \\
   - 2\beta _2 (t)\int\limits_0^\infty  {d\omega } g^2 (\omega )I(\omega ,t) \hfill \\
\end{gathered}
\end{equation}

where
\begin{equation}\label{A3}
I(\omega ,t) = \int\limits_0^t {} e^{ - i\left( {\omega  - \Omega
} \right)(t - t')} dt'
\end{equation}

Next, we change variables in (\ref{A3}), $t-t'=\tau$, and tend the
upper bound of integral to infinity.
\begin{equation}\label{A4}
\begin{gathered}
  I(\omega ,t) = \int\limits_0^t {dt'} e^{ - i\left( {\omega  - \Omega } \right)(t - t')}  = \int\limits_0^t {d\tau } e^{ - i\left( {\omega  - \Omega } \right)\tau }  \hfill \\
   \approx \int\limits_0^\infty  {d\tau } e^{ - i\left( {\omega  - \Omega } \right)\tau }  = \pi \delta (\omega  - \Omega ) - iP\frac{1}
{{\omega  - \Omega }} \hfill \\
\end{gathered}
\end{equation}

where $P\frac{1}{\omega-\Omega}$ denotes Cauchy's principal value.

Then, for (\ref{A1}), (\ref{A2}) we obtain:
\begin{equation}\label{A5}
\begin{gathered}
\frac{{d\beta _1 }}
{{dt}} =  - i\int\limits_0^\infty  {d\omega } g(\omega )\gamma _0 \left( \omega  \right)e^{ - i(\omega  - \Omega )t}  \hfill \\
   - 2\pi \beta _1 (t)g^2 (\Omega ) - 2\pi \beta _2 (t)g^2 (\Omega )\cos (k_\Omega  d) \hfill \\
   + i2\beta _1 (t)P\int\limits_0^\infty  {d\omega } \frac{{g^2 (\omega )}}
{{\omega  - \Omega }} + i2\beta _2 (t)P\int\limits_0^\infty
{d\omega } \frac{{g^2 (\omega )\cos (k_\omega  d)}}
{{\omega  - \Omega }} \hfill \\
\end{gathered}
\end{equation}

\begin{equation}\label{A6}
\begin{gathered}
  \frac{{d\beta _2 }}
{{dt}} =  - i\int\limits_0^\infty  {d\omega } g(\omega )\gamma _0 \left( \omega  \right)e^{ik_\omega  d} e^{ - i(\omega  - \Omega )t}  \hfill \\
   - 2\pi \beta _1 (t)g^2 (\Omega )\cos (k_\Omega  d) - 2\pi \beta _2 (t)g^2 (\Omega ) \hfill \\
   + i2\beta _1 (t)P\int\limits_0^\infty  {d\omega } \frac{{g^2 (\omega )\cos (k_\omega  d)}}
{{\omega  - \Omega }} + i2\beta _2 (t)P\int\limits_0^\infty
{d\omega } \frac{{g^2 (\omega )}}
{{\omega  - \Omega }} \hfill \\
\end{gathered}
\end{equation}

The quantity $P\int\limits_0^\infty  {d\omega } \frac{{g^2 (\omega
)}} {{\omega  - \Omega }}$ results in the shift of the qubit
frequency $\Omega$. We assume the shift is small and include it
implicitly in the definition of $\Omega$. As the coupling
$g(\omega)$ between qubit and the field is effective at the qubit
resonance, we take it off the Cauchy principal integral in
equations (\ref{A5}), (\ref{A6}) at the qubit  resonance
frequency. Then for the  Cauchy principal integral we obtain:
\begin{equation}\label{A7}
\begin{gathered}
  P\int\limits_0^\infty  {d\omega } \frac{{g^2 (\omega )\cos (k_\omega  d)}}
{{\omega  - \Omega }} = g^2 (\Omega )P\int\limits_0^\infty
{d\omega } \frac{{\cos (k_\omega  d)}}
{{\omega  - \Omega }} \hfill \\
   =  - \pi g^2 (\Omega )\sin (k_\Omega  d) \hfill \\
\end{gathered}
\end{equation}
For two-qubit system the rate of spontaneous emission can be found
from Fermi's golden rule:
\begin{equation}\label{A8}
 \Gamma  = 4\pi g^2 (\Omega )
\end{equation}

Combining the equations (\ref{A7}) and (\ref{A8}) with equations
(\ref{A5}) and (\ref{A6}) we obtain the equations (\ref{12}) and
(\ref{13}) which are given in the main text.

\section{Properties of sine and cosine integrals and some related integrals}

Here we use the conventional definitions for sine and cosine
integrals \cite{Grad2007}.
\begin{equation}\label{B1}
    \rm{si}(xy) =  - \int\limits_x^\infty  {\frac{{\sin zy}} {z}} dz
\end{equation}
\begin{equation}\label{B2}
    \rm{ci}(xy) =  - \int\limits_x^\infty  {\frac{{\cos zy}} {z}} dz
\end{equation}

\begin{equation}\label{B3}
\rm{Si}(xy) = \int\limits_0^x {\frac{{\sin zy}} {z}} dz
\end{equation}
\begin{equation}\label{B4}
\int\limits_0^\infty  {\frac{{\sin zy}} {z}} dz = \frac{\pi }
{2}sign(y)
\end{equation}
where $\rm{si}(xy)$ is defined on the whole real axis, while
$\rm{ci}(xy)$ is defined only for $x>0$.

Using these definitions it is not difficult to show that
\begin{equation}\label{B5}
\rm{si}(xy) = \rm{si}(xy) - \frac{\pi } {2}sign(y)
\end{equation}
\begin{equation}\label{B6}
\rm{si}( - xy) =  - \int\limits_{ - x}^\infty  {\frac{{\sin zy}}
{z}} dz =  - \rm{si}(xy) - \frac{\pi } {2}sign(y)
\end{equation}

Combining (\ref{B5}) and (\ref{B6}) we obtain a useful relation
\begin{equation}\label{B7}
    \rm{si}(xy)+\rm{si}(-xy)=-\pi sign(y)
\end{equation}

From definitions (\ref{B1}), (\ref{B2}), and (\ref{B3})  the
parity relations follow:
\begin{equation}\label{B8}
\begin{gathered}
    \rm{si}(x(-y))=-\rm{si}(xy);\,\rm{ci}(x(-y))=\rm{ci}(xy);\hfill\\ \rm{si}(x(-y))=-\rm{si}(xy)
\end{gathered}
\end{equation}
The exponential integral $E_1(z)$ in the expressions (\ref{40}),
(\ref{41}) is defined as follows \cite{Abram1964}:
\begin{equation}\label{E1}
    E_1(z) =  \int\limits_z^\infty  {\frac{{e^{-t }}} {t}} dt
\end{equation}

The behavior of scattered fields at large x and t follows from the
asymptote of the exponential integral function, sine, and cosine
integrals \cite{Grad2007, Jahnke}:

\begin{equation}\label{Asymp1}
    \rm{si}(x)\approx -\frac{\cos(x)}{x}-\frac{\sin(x)}{x^2};
    \quad \rm{ci}(x)\approx \frac{\sin(x)}{x}-\frac{\cos(x)}{x^2}
\end{equation}
where $x\gg 1$.
\begin{equation}\label{Asymp2}
    E_1(z)\approx \frac{e^{-z}}{z}\left(1-\frac{1}{z}\right)
\end{equation}
where $|z|\gg 1$.

Below we illustrate the application of above formulae for the
calculation of some integrals which we use throughout the paper.
\begin{equation}\label{B9}
I_2 (x,t) = \int\limits_0^\infty  {d\omega } \frac{{e^{i(\omega  -
\omega _S )t}  - 1}} {{\omega  - \omega _S }}e^{i\frac{\omega }
{{v_g }}\left( {x - v_g t} \right)}
\end{equation}
This integral, where $x>0$ and $x-v_gt<0$ describes the forward
travelling wave between qubits $0<x<d$ as well as behind the
second qubit, $x>d$.

Changing the variables in the integrand of (\ref{B9}),
$z=\omega-\omega_S$, $\tau=x/v_g$, $T=(x-v_gt)/v_g$ we obtain
\begin{equation}\label{B10}
I_2 (x,t) = e^{i\omega _S T} \left( {\int\limits_{ - \omega _S
}^\infty  {dz} \frac{{e^{iz\tau } }} {z} - \int\limits_{ - \omega
_S }^\infty  {dz} \frac{{e^{izT} }} {z}} \right)
\end{equation}
The calculation of the first integral in (\ref{B10}) yields:
\begin{equation}\label{B11}
\begin{gathered}
  \int\limits_{ - \omega _S }^\infty  {dz} \frac{{e^{iz\tau } }}
{z} = \int\limits_{ - \omega _S }^\infty  {dz} \frac{{\cos z\tau
}} {z} + i\int\limits_{ - \omega _S }^\infty  {dz} \frac{{\sin
z\tau }}
{z} \hfill \\
   = \int\limits_{ - \omega _S }^{\omega _S } {dz} \frac{{\cos z\tau }}
{z} + \int\limits_{\omega _S }^\infty  {dz} \frac{{\cos z\tau }}
{z} + i\int\limits_{ - \omega _S }^\infty  {dz} \frac{{\sin z\tau
}}
{z} \hfill \\
   =  - \rm{ci}(\omega _S \tau ) - \emph{i}\,\rm{si}( - \omega _S \tau ) \hfill \\
\end{gathered}
\end{equation}
Similar expression we obtain for second integral in (\ref{B10}):
\begin{equation}\label{B12}
\int\limits_{ - \omega _S }^\infty  {dz} \frac{{e^{izT} }} {z} = -
\rm{ci}(\omega _S T) - \emph{i}\,\rm{si}( - \omega _S T)
\end{equation}
 Therefore, for $I_2(x,t)$ we obtain:
 \begin{equation}\label{B13}
 \begin{gathered}
    I_2 (x,t) = e^{i\omega _S
    T}\hfill\\\times\left(-\rm{ci}(\omega_S\tau)-\emph{i}\,\rm{si}(-\omega_S\tau)+\rm{ci}(\omega_ST)+\emph{i}\,\rm{si}(-\omega_ST)\right)
\end{gathered}
\end{equation}
Using the relation (\ref{B7}) we rewrite (\ref{B13}) as follows:
\begin{equation}\label{B14}
 \begin{gathered}
    I_2 (x,t) = e^{i\omega _S
    T}\hfill\\\times\left(-\rm{ci}(\omega_S\tau)+\emph{i}\,\rm{si}(\omega_S\tau)+\rm{ci}(\omega_ST)-\emph{i}\,\rm{si}(\omega_ST)\right)\hfill\\
    +i\pi e^{i\omega _S
    T}\left( sign(\tau)- sign(T)\right)
\end{gathered}
\end{equation}
Here $\tau>0$, and $T<0$. Therefore, we obtain:

\begin{equation}\label{B15}
 \begin{gathered}
    I_2 (x,t) = e^{i\omega _S
    T}\hfill\\\times\left(-\rm{ci}(\omega_S\tau)+\emph{i}\,\rm{si}(\omega_S\tau)+\rm{ci}(\omega_S|T|)+\emph{i}\,\rm{si}(\omega_S|T|)\right)\hfill\\
    +2i\pi e^{i\omega _ST}
\end{gathered}
\end{equation}
where $\tau=x/v_g>0$, $T=(x-v_gt)/v_g<0$.

For integral $I_2(x-d,t)$ which describes the forward travelling
wave between qubits, $x-d<0$ we obtain from (\ref{B14}), where $x$
is replaced with $x-d$:
\begin{equation}\label{B16}
 \begin{gathered}
    I_2 (x-d,t) = e^{i\omega _S
    T}\hfill\\\times\left(-\rm{ci}(\omega_S|\tau|)-\emph{i}\,\rm{si}(\omega_S|\tau|)+\rm{ci}(\omega_S|T|)+\emph{i}\,\rm{si}(\omega_S|T|)\right)
\end{gathered}
\end{equation}
where $\tau=(x-d)/v_g<0$, $T=(x-d-v_gt)/v_g<0$.

Next, we consider the integral (\ref{60}) which describes the
backward travelling wave in front of the first qubit $x<0$.

\begin{equation}\label{B17}
J_2 (x,t) = \int\limits_0^\infty  {d\omega } \frac{{e^{i(\omega  -
\omega _S )t}  - 1}} {{\omega  - \omega _S }}e^{ - i\frac{\omega }
{{v_g }}\left( {x + v_g t} \right)}
\end{equation}
where $x<0$, $x+v_gt>0$.

The calculation of this integral is similar to that of $I_2(x,t)$.
For $J_2(x,t$ we obtain the following result:
\begin{equation}\label{B18}
 \begin{gathered}
    J_2 (x,t) = e^{-i\omega _S
    T}\hfill\\\times\left(-\rm{ci}(\omega_S\tau)-\emph{i}\,\rm{si}(\omega_S\tau)+\rm{ci}(\omega_ST)+\emph{i}\,\rm{si}(\omega_ST)\right)\hfill\\
    -i\pi e^{-i\omega _S
    T}\left( sign(\tau)- sign(T)\right)
\end{gathered}
\end{equation}
 Therefore, for $J_2(x,t)$ we
finally obtain:
\begin{equation}\label{B19}
 \begin{gathered}
    J_2 (x,t) = e^{-i\omega _S
    T}\hfill\\\times\left(-\rm{ci}(\omega_S|\tau|)+\emph{i}\,\rm{si}(\omega_S|\tau|)+\rm{ci}(\omega_ST)+\emph{i}\,\rm{si}(\omega_ST)\right)\hfill\\
    +2i\pi e^{-i\omega _S T}
\end{gathered}
\end{equation}
where $\tau=x/v_g<0$, $T=(x+v_gt)/v_g>0$.

There are two integrals which describe the backward travelling
wave between qubits, $J_2(x,t)$ where $x>0$, and $J_2(x-d),t$
where $x-d<0$. The quantity $J_2(x,t)$ with $x>0$ follows from
(\ref{B18}) where $\tau=x/v_g>0$:
\begin{equation}\label{B20}
\begin{gathered}
 J_2 (x,t) = e^{-i\omega _S
    T}\hfill\\\times\left(-\rm{ci}(\omega_S\tau)-\emph{i}\,\rm{si}(\omega_S\tau)+\rm{ci}(\omega_ST)+\emph{i}\,\rm{si}(\omega_ST)\right)
\end{gathered}
\end{equation}
where $\tau=x/v_g>0$, $T=(x+v_gt)/v_g>0$.

The integral $J_2(x-d,t)$ is obtained from equation (\ref{B19})
where $x$ is replaced with $x-d$, and where $\tau=(x-d)/v_g<0$,
$T=(x-d+v_gt)/v_g>0$.

\end{document}